\documentclass[aps,twocolumn,showpacs,bibnotes]{revtex4-1}  

\usepackage{graphicx}  
\usepackage{dcolumn}   
\usepackage{bm}        
\usepackage{amssymb}   
\usepackage{amsmath}
\usepackage{bbm,amsfonts}

\usepackage{color}
\usepackage{amsthm}
\usepackage{setspace}
\usepackage{nomencl}
\usepackage{amstext} 
\usepackage{subfigure}

\usepackage{tabularx}
\usepackage[utf8]{inputenc}

\usepackage[colorlinks=true]{hyperref}

\newcommand{\Eop}{\mathcal{E}}

\begin{document}

\title{Reducing unitary and spectator errors in cross resonance with optimized rotary echoes}
\author{Neereja Sundaresan}\thanks{These authors have contributed equally to this work}\affiliation{IBM Quantum, T. J. Watson Research Center, Yorktown Heights, NY 10598}
\author{Isaac Lauer}\thanks{These authors have contributed equally to this work}\affiliation{IBM Quantum, T. J. Watson Research Center, Yorktown Heights, NY 10598}
\author{Emily Pritchett}\thanks{These authors have contributed equally to this work}\affiliation{IBM Quantum, T. J. Watson Research Center, Yorktown Heights, NY 10598}
\author{Easwar Magesan}\thanks{These authors have contributed equally to this work}\affiliation{IBM Quantum, T. J. Watson Research Center, Yorktown Heights, NY 10598}
\author{\\Petar Jurcevic}\affiliation{IBM Quantum, T. J. Watson Research Center, Yorktown Heights, NY 10598}
\author{Jay M. Gambetta}\affiliation{IBM Quantum, T. J. Watson Research Center, Yorktown Heights, NY 10598} 
\date{July 2020}

\begin{abstract}
   We present an improvement to the cross resonance gate realized with the addition of resonant, target rotary pulses. These pulses, applied directly to the target qubit, are simultaneous to and in phase with the echoed cross resonance pulses. Using specialized Hamiltonian error amplifying tomography, we confirm a reduction of error terms with target rotary -- directly translating to improved two-qubit gate fidelity. Beyond improvement in the control-target subspace, the target rotary reduces entanglement between target and target spectators caused by residual quantum interactions. We further characterize multi-qubit performance improvement enabled by target rotary pulsing using unitarity benchmarking and quantum volume measurements, achieving a new record quantum volume for a superconducting qubit system. 
\end{abstract}

\maketitle

\section{Introduction}
Improving the performance of near-term quantum systems is important to quantum information technology today.  These systems, realized and made available for general use in recent years,
demonstrate  multi-partite entanglement, algorithms, and fault-tolerant protocols~\cite{Riste2015,Takita2017,Kandala2017,Hempel2018,Havlicek2019,Brydges2019,Nam2020,Wei2020}. However, much work remains before we can execute quantum circuits with sufficient fidelity to  consistently demonstrate quantum advantage. Here we show improvements in our understanding and in the performance of our two-qubit gates in multi-qubit systems which provide a significant increase in various performance metrics including the quantum volume (QV)~\cite{Cross2019} of the system. \\

The cross resonance (CR) gate has emerged as a promising two-qubit entangling gate in superconducting quantum computing architectures ~\cite{Paraoanu2006,Rigetti2010,Chow2011}. The CR gate generates entanglement using microwave pulses without the need for qubit or coupling tunability, which simplifies scaling to larger numbers of qubits by reducing the number of input lines and overhead of control electronics. In this respect, the CR gate compares favorably to those approaches which require variable magnetic flux to tune qubit frequencies~\cite{DiCarlo2009,Bialczak2010,Barends2014} and/or bus couplings~\cite{McKay2016}, or use microwave pulses applied to buses to induce entanglement~\cite{Paik2016}.  \\

While CR pulses enhance the entangling interaction between two qubits, they introduce unwanted errors for implementing high-fidelity gates~\cite{Sheldon2016,Magesan2020,Tripathi2019,Malekakhlagh2020}. A simple echo sequence corrects most of the unwanted terms in the CR Hamiltonian leading to significant increase in the two-qubit gate fidelity~\cite{Corcoles2013}. Errors can be further reduced by proper choice of calibration frame~\cite{McKay2019} and devising strategies to negate effects from classical crosstalk~\cite{Sheldon2016}.
As qubit coherence improves, however, we find higher-order unitary errors such as those arising from always-on ZZ interactions limit gate fidelity below the level set by coherence. \\ 

The effects of static $ZZ$ coupling go beyond creating errors in the two-qubit subspace. Interactions with other qubits in the device  -- ``spectators'' -- cause unwanted entanglement to accumulate across the system.   
While these ZZ-induced errors can be corrected with more complicated pulse sequences, such as higher-order echoes to address spectator-induced error ~\cite{Takita2016} or additional single-qubit rotations to correct unitary errors in the two-qubit subspace, we show here that a resonant drive of the target qubit reduces both types of error simultaneously without increasing the duration or the depth of the two-qubit gate sequence. 
This ``target rotary'' pulsing, presented schematically in Fig. ~\ref{figure:Fig1}(a),  is performed in parallel to the CR drive of the control qubit and also switches sign in the standard two-pulse echo sequence. \\

We develop specialized tomographic error amplification sequences to measure how target rotary pulses reduce unitary errors.  These errors are comprised of terms in the control-target subspace and as well as terms containing entanglement between the target qubit and target spectator qubits.  We verify the impact of the rotary tone in reducing these errors with both randomized benchmarking (RB)~\cite{Knill08,Magesan2011} and benchmarking the noise unitarity via purity measurements~\cite{Wallman_2015}. We also find that the QV, a holistic measure of device performance affected by unitary and purity errors amongst others, increases to 32 with the addition of target rotary on many 5-qubit subsystems of the 20-qubit device tested.


\section{Gate Errors}

\subsection{Errors in the two-qubit subspace}

Consider a system of two detuned transmons coupled by a dipole interaction, which can be modeled as two Duffing oscillators with a Jaynes-Cummings interaction,
\begin{eqnarray}
H = \sum_{j=0}^1\bigg[~ \omega_j b_j^\dagger b_j &+& \frac{\delta_j}{2}b_j^\dagger b_j \left(b_j^\dagger b_j - \mathbbm{1}\right)\bigg] \\
&+& ~J \left(b_0^\dagger b_1 + b_0 b_1^\dagger\right)\nonumber
\end{eqnarray} 
where $\omega_j$ and $\delta_j$ are the transmon frequencies and anharmonicities respectively, $J$ is the transmon-transmon exchange coupling, $b_j$ is the lowering operator for the $j^\mathrm{th}$ qubit, $\mathbbm{1}$ is the identity operator, and for simplicity we have taken $\hbar=1$. Diagonalizing the Hamiltonian $H$ produces a static $ZZ$ interaction in the resulting dressed basis~\cite{DiCarlo2009}. A CR tone, which drives one transmon (designated the ``control") at the dressed frequency of the other (designated the ``target"),  produces an effective Hamiltonian of the form~\cite{Magesan2020},
\begin{gather}
H (\Omega)= \nu_{ZX}\frac{ZX}{2} + \nu_{IZ}\frac{IZ}{2}
+  \nu_{IX}\frac{IX}{2} \nonumber \\
+\nu_{ZI}\frac{ZI}{2} +  \nu_{ZZ}\frac{ZZ}{2},\label{eq:CRHam}
\end{gather}
where we choose the convention of control being first and target second in the tensor product reading from left to right. There are various terms in the effective Hamiltonian of Eq.~\ref{eq:CRHam}, including a $ZX$ conditional rotation. We would like to isolate the $ZX$ term so that the evolved unitary is $ZX_{\frac{\pi}{2}}=e^{-i\frac{\pi}{4}ZX}$, which is locally equivalent to the standard CNOT gate that easily compiles into universal quantum circuits.  Our goal here is to identify the unwanted CR Hamiltonian error terms that remain after standard echo sequences. Later we will devise strategies to characterize these terms through error amplification sequences and mitigate their effects through additional target rotary tones.\\

\subsubsection{Origin of unwanted errors on the target qubit:  IY and IZ}

\begin{figure}
\includegraphics[width=8.7cm]{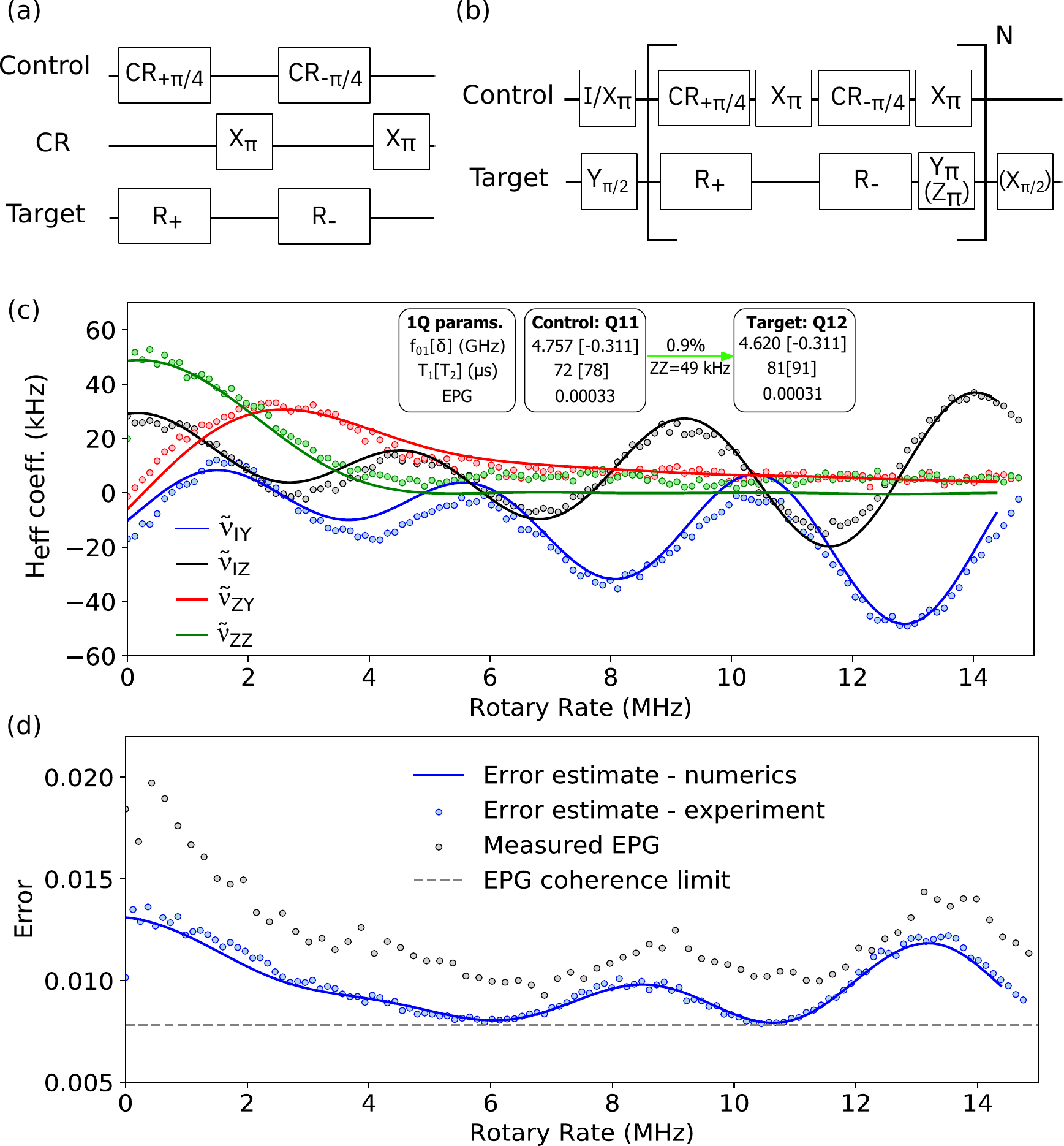}
\caption{(a) Diagram of the echoed cross resonance (CR) with target rotary pulse sequence, where $R_{\pm}$ denotes a fixed $\pm X$ rotation on target. (b) Hamiltonian Error Amplifying Tomography (HEAT): example of four gate sequences used to accurately reconstruct echoed CR Hamiltonian error terms.  (c) Hamiltonian errors with target rotary  on a pair of qubits (inset).  Solid lines are numerical fits to the data described in the Appendix. (d) Two-qubit EPG (black circles) track the error in blue estimated by only considering the contribution from the four error terms measured in (c) added to the coherence limit set by the measured $T_1$s and $T_2$s for a gate time of 484 ns (dashed line).}
\label{figure:Fig1}
\end{figure}

The coefficients $\nu_{ij}$ of Eq.~\ref{eq:CRHam} are a function of the system parameters and CR drive amplitude $\Omega$~\cite{Magesan2020}. The diagonal coefficients, $\nu_{IZ}$, $\nu_{ZI}$, and $\nu_{ZZ}$ are even order in $\Omega$ while the non-diagonal coefficients $\nu_{IX}$ and $\nu_{ZX}$ are odd order in $\Omega$. Hence reversing the sign of $\Omega$ only reverses the sign of the non-diagonal coefficients. The standard  two-pulse echo sequence, as shown in Fig. \ref{figure:Fig1}(a) but without the target pulses, removes most of the unwanted terms from Eq.~\ref{eq:CRHam}. Defining $H_{\rm 1}\equiv H(\Omega)$ and ${H}_{\rm 2}\equiv XI \cdot H(-\Omega)\cdot XI$ to be the system Hamiltonians for the positive tone and the rotated CR tone with negative drive amplitude respectively,
\begin{equation}
H_{\rm eff}=\frac{i}{\tau}{\rm ln}\bigg[e^{-iH_2t}e^{-iH_1t}\bigg],
\end{equation}
describes the effective Hamiltonian of the entire pulse sequence. 
$H_1$ and $H_2$ are the same up to a sign change on the coefficients of $IX$, $ZI$, and $ZZ$, and $H_{\rm eff}$
is predominantly a $ZX$ term where $\tau$ is the effective time satisfying $H_{\rm eff}\tau\simeq \frac{\pi}{4}ZX$. We would like to identify and reduce the other terms present in $H_{\rm eff}$.  \\

Let $U= e^{-iH_{\rm eff}\tau} = U_1\cdot U_2$ be the unitary evolution of the entire echo sequence, where $U_1$ and $U_2$ are the unitary operators corresponding to $H_1$ and $H_2$ respectively. We calculate the Pauli coefficients of $U$ in Appendix~\ref{sec:Echo}. The terms that appear with the same signs in  $H_1$ as in $H_2$, specifically $II$, $IZ$, and $ZX$, are nonzero in $U$. Interestingly, the $IY$ coefficient of $U$ is also non-zero and second order in the $\nu_{ij}$ since the pairs $(IX,IZ)$ and $(ZX,ZZ)$ are each composed of anti-commuting operators that have opposite signs in $H_1$ and $H_2$. $H_{\rm eff}$ can therefore be expanded in the Pauli basis as,
\begin{eqnarray}
    H_{\rm eff}
    &=&\tilde{\nu}_{ZX}\frac{ZX}{2} + H_{\rm err},
\end{eqnarray} 
where
\begin{eqnarray}
    H_{\rm err} \equiv \tilde{\nu}_{IY}\frac{IY}{2} + \tilde{\nu}_{IZ}\frac{IZ}{2},
\end{eqnarray} 
and $\tilde{\nu}_{ij}$ are used to denote the Pauli coefficients of the effective Hamiltonian of the entire pulse sequence. The exact expressions for the $\tilde{\nu}_{ij}$ are given as a function of the Pauli coefficients of only the positive CR pulse, which are denoted $\nu_{ij}$ (see Appendix~\ref{sec:Echo}). Note the coefficients of $U$ are simply proportional to the coefficients of $H_{\rm eff}$ when $\tilde{\nu}_{ZX}\tau=\frac{\pi}{2}$. Nonzero $\tilde{\nu}_{IY}$ and $\tilde{\nu}_{IZ}$ are the source of unitary errors within the 2Q subspace of the CR gate, while any corrections to $\nu_{ZX}$ can be absorbed into the CR gate calibration.
Note that $\tilde{\nu}_{IY}$ errors caused by static $ZZ$ contribute at second order even with a perfectly executed echo sequence and no crosstalk.\\

\subsubsection{Rotary Pulsing without Crosstalk}

The addition of a rotary tone applied in parallel to the CR tone, as shown in Fig.~\ref{figure:Fig1}(a), can be beneficial for modifying the size of terms in the effective Hamiltonian of the entire pulse sequence while not introducing new error types or extending the gate time (see Appendix~\ref{sec:Rotary_echo}). The main effect of the rotary pulse on just the CR + rotary tone is to tune $\nu_{IX}$, however the off-resonant driving of higher levels of the target also modifies both $\nu_{IZ}$ and $\nu_{ZZ}$. 

In the case of no classical crosstalk, as shown in Appendix~\ref{sec:Echo_ct_rotary}, the rotary tone can be used to eliminate $\tilde{\nu}_{IY}$. To understand why this is the case, note that
\begin{equation}
\tilde{\nu}_{IY} = \frac{\pi}{\sqrt{2}t}\frac{\chi_0}{\eta_+ \eta_-}\sin{\frac{\eta_+ t}{2}}\sin{\frac{\eta_- t}{2}}
\end{equation}
for pulse duration $t$ where the coefficient
\begin{equation}
\chi_0 \equiv \nu_{IX}\nu_{IZ}-\nu_{ZX}\nu_{ZZ}
\end{equation}
depends on the relative size of the noncommuting pairs $(IX,IZ)$ and $(ZX,ZZ)$ in the CR Hamiltonian, and
\begin{equation}
\eta_\pm^2 \equiv (\nu_{IX} \pm \nu_{ZX})^2+(\nu_{IY} \pm \nu_{ZY})^2+(\nu_{IZ} \pm \nu_{ZZ})^2.
\end{equation}
The full set of conditions for which $\tilde{\nu}_{IY}$ can be set to 0 are given in Appendix~\ref{sec:Echo_ct_rotary}. As an approximation note that realistically $\nu_{IX}$ and $\nu_{ZX}$ are much larger in magnitude than all other coefficients, allowing us to write
\begin{equation}
\tilde{\nu}_{IY}\simeq \frac{\pi\chi_0}{\sqrt{2}t} \left(\frac{\cos{\nu_{IX}t}-\cos{\nu_{ZX} t}}{\nu_{IX}^2+\nu_{ZX}^2}\right).
\end{equation}
We see that $\tilde{\nu}_{IY}$  oscillates with the increasing magnitude of $\nu_{IX}$. The CR tone amplitude and $t$ are fixed in order to implement a $ZX_{\frac{\pi}{2}}$ gate and $\nu_{ZX}$ is insensitive to the rotary amplitude.    Within this approximation, $\tilde{\nu}_{IY}=0$ when one of conditions
\begin{enumerate}
    \item $\nu_{IX}\nu_{IZ}=\nu_{ZX}\nu_{ZZ}$
    \item $\nu_{IX}=\pm \nu_{ZX}+n\omega, \: n\in\mathbb{Z}  \ \& \ n \neq 0$
\end{enumerate}
are met, where $\omega=\frac{2\pi}{t}$. Therefore as the rotary amplitude is swept, $\tilde{\nu}_{IY}=0$ when $\nu_{IX} \approx \nu_{ZX}+n\omega$ for different values of $n\neq 0$. Since no additional error-types are created in the echoed gate, only $\tilde{\nu}_{IZ}$ is present in the error Hamiltonian, which can be corrected by updating the frame of the target qubit. As the rotary amplitude is swept there are potentially multiple solutions to $\tilde{\nu}_{IY}=0$, allowing it to be constrained by other considerations such as minimizing classical crosstalk or spectator effects.  While these arguments allow us to understand the effect of target rotary pulsing intuitively, a quantitative model of experiment includes classical crosstalk.  \\

\subsubsection{Rotary Pulsing with Crosstalk} 
Classical crosstalk, arising from some amount of the CR drive inadvertently reaching the target qubit, is an important potential source of error~\cite{Sheldon2016}. In this case the CR Hamiltonian in Eq.~\ref{eq:CRHam} can have non-zero $\nu_{IY}$ and  $\nu_{ZY}$ that depend on the CR drive amplitude and the phase of the signal seen by the target, which is shifted from the phase seen by the control by an amount that depends in detail on circuit layout.
The crosstalk and rotary tones have the same frequency with different amplitudes and phases. Taken together they create a single tone of the same frequency with a phase and amplitude modulated by the phases and amplitudes of the individual tones. 
The inclusion of crosstalk from the CR tone causes the rotary echo Hamiltonian to have nonzero $\tilde{\nu}_{IX}$, $\tilde{\nu}_{ZY}$, and $\tilde{\nu}_{ZZ}$ in addition to $\tilde{\nu}_{IY}$ and $\tilde{\nu}_{IZ}$.
 Analytic expressions for the error terms with classical crosstalk are given in Appendix~\ref{sec:Echo_ct_rotary}. Importantly, we find that as the rotary tone amplitude increases, the sizes of $\tilde{\nu}_{IX}$, $\tilde{\nu}_{ZY}$, and $\tilde{\nu}_{ZZ}$ quickly reduce. 

\subsubsection{Hamiltonian Reconstruction}
We introduce tailored tomographic methods -- which we refer to as Hamiltonian Error Amplifying Tomography (HEAT) -- that amplify and measure errors in our system. The HEAT sequences, shown in Fig. \ref{figure:Fig1}(b) (Appendix~\ref{sec:HEAT_re} for details), can be used in scenarios where errors are of the form $Z\otimes A$ or $I\otimes A$ with $A\in X,Y,Z$. Both analytics and numerical analyses show that these are the main errors for current CR gate implementations.  In the case of the rotary echo CR gate, we have that $\tilde{\nu}_{ZX}\tau \simeq \frac{\pi}{2}$ and the HEAT sequences are sensitive to the small $\tilde{\nu}_{IY}$ and $\tilde{\nu}_{IZ}$ errors.  \\

HEAT can be compared favorably to more traditional direct Hamiltonian tomography sequences~\cite{Sheldon2016} that apply in more generality, but require more data collection and fitting of time series data to estimate the effective Hamiltonians of individual pulses. In addition, small error terms can be challenging to detect with these methods.  By comparison HEAT uses repetition to amplify known errors to the completed echo sequence, easing their detection, and allowing us to calibrate the target rotary pulse magnitude by directly minimizing these errors.\\

We now test these concepts on the 20-qubit  \emph{$ibmq\_johannesburg$} system \footnote{\emph{$ibmq\_johannesburg$} v1.0.2, IBM Quantum team. Retrieved from https://quantum-computing.ibm.com (2020)}, illustrated in more detail in Appendix~\ref{sec:Volume}. Focusing on qubits Q11 and Q12, we show the results of the HEAT sequence in Fig. \ref{figure:Fig1}(c),
tracking the dependence of the error terms on target rotary rate.  The data fits well to a numerical model (solid curves), outlined in Appendix~\ref{sec:Numerics}, that allows for classical crosstalk. This crosstalk introduces a large $ZZ$ error (in addition to that corrected by the echo sequence) that quickly decays with increasing rotary amplitude.   The measured Hamiltonian errors translate directly into gate error, as shown in Fig. \ref{figure:Fig1}(d), where the total error in \ref{figure:Fig1}(c) is related to that measured by 2QRB. In this case, adding target rotary reduced the 2Q error per gate by nearly a full percentage point, bringing gate error significantly closer to the limit set by coherence.

\begin{figure}
\includegraphics[width=8.6cm]{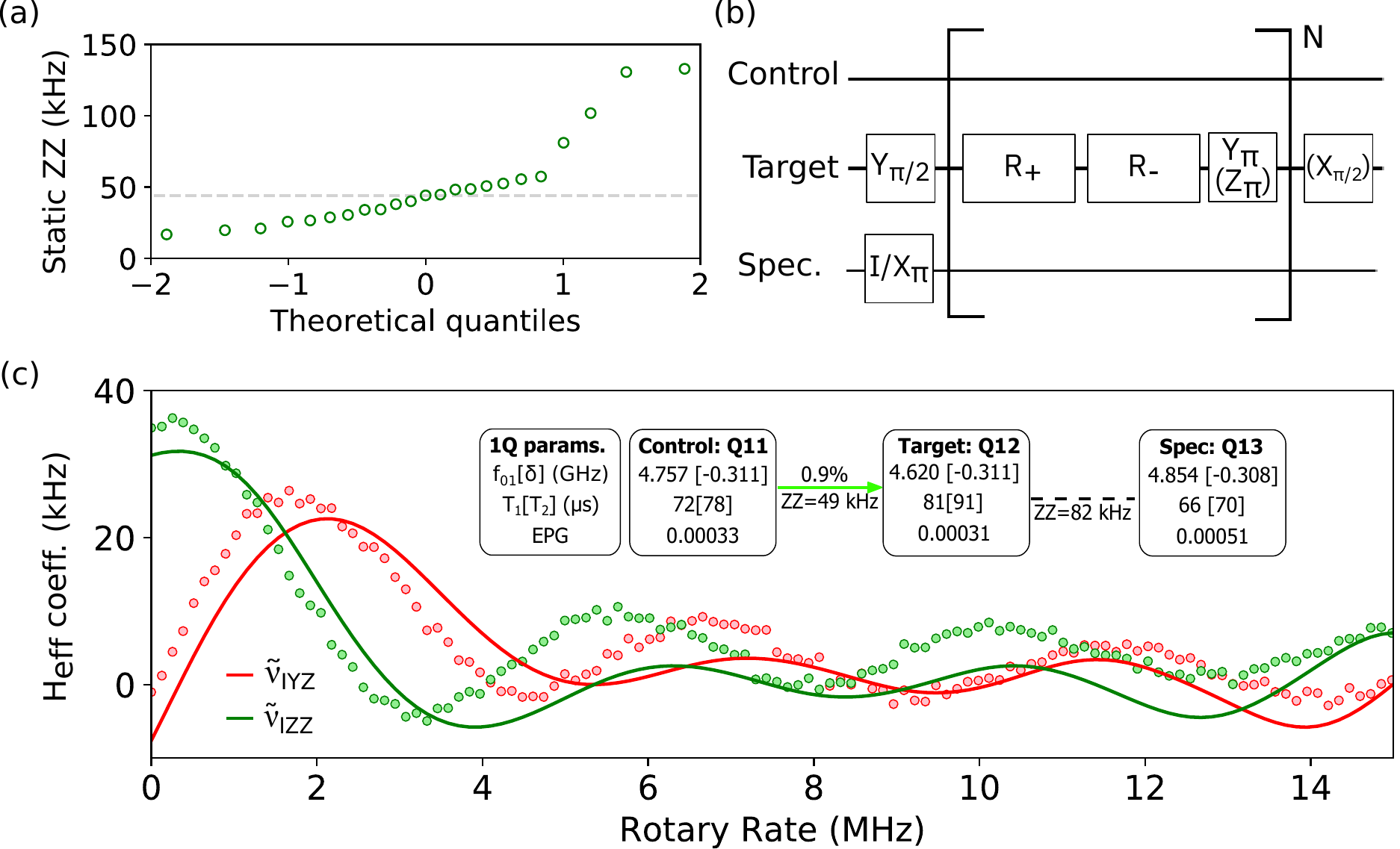}
\caption{(a) Quantile distribution of measured static $ZZ$ between coupled qubits on \emph{$ibmq\_johannesburg$}. Median ZZ denoted by gray dashed line. (b) Modified HEAT sequence to measure dominant target-target spectator entangling Hamiltonian terms. (c) Hamiltonian errors with target rotary (numerical fits (solid line) and experiment (circles)) on a pair of qubits (inset). }
\label{figure:Fig2}
\end{figure}

\subsection{Spectator Errors}
In addition to the errors within the two-qubit control-target subspace, coherent errors due to coupling with other nearest-neighbor qubits, or ``spectators" are known to be significant~\cite{Takita2016}. In our architecture, the impact of spectators mainly arises from the static $ZZ$ coupling between connected qubits. Due to the distribution of qubit frequencies, most qubit pairs on \emph{$ibmq\_johannesburg$} have static $ZZ$ below 50 kHz, with a few worse pairs exhibiting greater than 100 kHz $ZZ$ coupling as shown in Fig. \ref{figure:Fig2}(a).  In multi-qubit devices, we encounter spectators on both the control and target qubits, but the standard two-pulse echo sequence is only effective at reducing errors caused by control spectators (qubits that couple significantly only to the control qubit). As such, it is important to find methods that provide first-order correction to static $ZZ$ errors arising from target spectators (qubits that couple significantly only to the target qubit).\\ 
 
 We find that in addition to reducing unitary errors in the 2Q subspace, the target rotary pulse plays a critical role in reducing undesired target-target spectator entanglement errors. For clarity we denote Pauli rotations in the three qubit subspace with Control~$\otimes$~Target$~\otimes~$Target  Spectator ordering. As the target-target spectator dynamics are dominated by the target rotary pulse (the CR pulse has no significant effect on a target spectator), we utilize the HEAT sequences (shown in Fig. \ref{figure:Fig2}(b)) to identify and track the dependence of dominant entangling Hamiltonian terms $IYZ$ and $IZZ$ with target rotary  amplitude (see Appendix~\ref{sec:HEAT_st} for details).  The results, shown in Fig. \ref{figure:Fig2}(c), reveal rotary amplitudes where  both $IYZ$ and $IZZ$ are reduced. \\

In order to understand the reduction in $IYZ$ and $IZZ$ with target rotary amplitude we develop an effective Hamiltonian model for the target- target spectator system. The initial system Hamiltonian is modeled as 
\begin{align}
H&=\sum_{j=0}^1{\omega_j}b_j^\dagger b_j + \frac{\delta_j}{2}b_j^\dagger b_j\left(b_j^\dagger b_j-\mathbbm{1}\right) \nonumber \\ 
&+ J\left(b_0^\dagger b_1+b_0b_1^\dagger\right) + \Omega \cos(\omega_d t)\left(b_1^\dagger + b_1\right).
\end{align}
Diagonalizing the time-independent part of the Hamiltonian gives the dressed frequencies $\tilde{\omega}_j$ as well as the static $ZZ$ coefficient $\xi=\frac{J^2\left(\delta_1+\delta_2\right)}{\left(\Delta+\delta_1\right)\left(\Delta-\delta_2\right)}$. Following a similar procedure to~\cite{Magesan2020} gives to first-order in $\Omega$ an effective Hamiltonian $H^{(1)}$ with non-zero $ZZ$ coefficient $\tilde{\nu}_{ZZ}=\xi$ as well as non-diagonal terms including $\Omega$. Going to higher orders in $\Omega$ produces non-zero $IZ$ and $ZI$ coefficients as well as shifts on the off-diagonals. Switching the sign of $\Omega$ only changes the sign of the off-diagonal elements. Defining $H^{(1)}_{\pm} = H^{(1)}(\pm\Omega)$, $R_{\pm}= e^{-iH^{(1)}_{\pm} t}$, and $R=R_-R_+$, we find to first order in $\xi$ that the effective generating Hamiltonian has coefficients,
\begin{align}
\tilde{\nu}_{YZ}&\approx \frac{\xi (1-\cos (\Omega t))}{\Omega t},\nonumber \\
\tilde{\nu}_{ZZ}&\approx \frac{\xi \sin (\Omega t)}{\Omega t},
\end{align}
which decay as $\frac{1}{\Omega}$ with a scale set by $\xi$.

To further characterize the entangling error originating from target-target spectator coupling, we perform purity randomized benchmarking sequences and extract the unitarity of the noise via the decay rate~\cite{Wallman_2015,Feng2016}.  For a general quantum operation $\Eop$ the unitarity, denoted $u_\Eop$, is given by
\begin{align}
u_\Eop&= \frac{1}{d^2-1}\text{tr}\left([\Eop_*]^\dagger[\Eop_*]\right),
\end{align}
where $d$ is the dimension of the system, $[\Eop]$ is taken to be the representation of $\Eop$ with respect to the orthonormal Pauli basis, and $[\Eop_*]$ is the unital part of $[\Eop]$. Suppose the total system is comprised of $n$ subsystems each of dimension $d_j$ with local unital noise $\Eop_j$ acting on each subsystem. As shown in Appendix~\ref{sec:PurityRB}, the unitarity of the total system, $u_\Eop$, is given by $u_\Eop^{\rm p}$ where $ u_\Eop^{\rm p}$ is related to the unitarity of the individual systems via
\begin{align}
u_\Eop^{\rm p}&=\frac{1}{d^2-1}\left(\Pi_{j=1}^n\left(1+(d_j^2-1)u_{\Eop_j}\right)-1\right).
\end{align}
$u_\Eop = u_\Eop^{\rm p}$ is a very good approximation when including small non-unital effects. In the case of $d_1=4$, $d_2=2$, and including $T_1$ effects we find for the C+T-S system
\begin{align}\label{eq:unitprod}
u_\Eop^{\rm p} &= \frac{1}{63}\Big[ 15 u_{\Eop_1}\left(1+\gamma_{a,3}^2\right) + 45u_{\Eop_1}u_{\Eop_2} \nonumber \\
&+ 3u_{\Eop_2}(1+\gamma_{a,1}^2)(1+\gamma_{a,2}^2)\Big],
\end{align}
where $\gamma_{a,j}$ is the $j$'th decay probability.
This product unitarity being equal to the full three qubit unitarity implies that the subsystems are not entangled.\\ 

\begin{figure}
\includegraphics[scale=0.65]{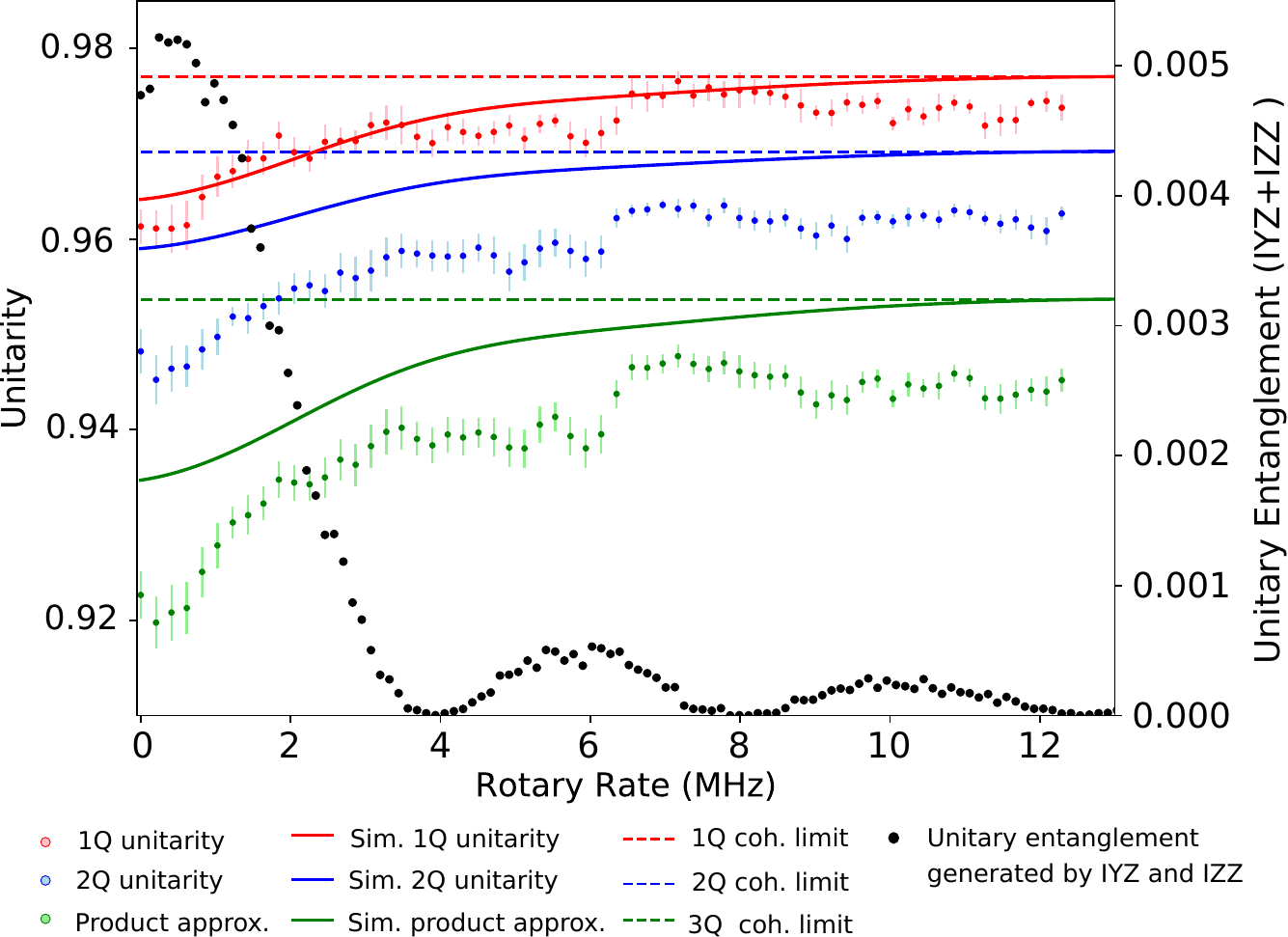}
\caption{Unitarity of the target spectator (red) and control-target (blue) subspaces extracted from purity RB on each subspace while performing Clifford gates in the opposite subspace. The product unitarity (Eq.~\ref{eq:unitprod} in the text) is shown in green, as well as the estimated unitarity in each case including only measured $T_1$s and $T_2$s (dashed lines) and also unitary errors from fits above (solid curves). Error bars are standard deviation of measurements sampled over approximately 1 week to represent drift in unitarities. Black dots (right axis) show the unitary entanglement contributed by the $\tilde{\nu}_{IYZ}$ and $\tilde{\nu}_{IZZ}$ measured in Fig. \ref{figure:Fig2} over the same range of rotary amplitudes.}
\label{figure:Fig3}
\end{figure}

To estimate the unitarity of the constituent control-target (2Q) and target spectator (1Q) subsystems of our 3Q system, we perform purity 
RB on the 2Q subsystem while driving 1Q Cliffords on the spectator, and vice versa.  To illustrate the worst case scenario, the spectator is idled during each 2Q gate to prevent refocusing of spectator errors originating from static ZZ during the CR. The results as a function of rotary amplitude are shown in Fig. \ref{figure:Fig3}.  We find a strong reduction in unitarity error per Clifford in both the 2Q and 1Q subsystems with the addition of rotary, consistent with the reduction of entangling error terms identified in Fig. ~\ref{figure:Fig2}(c).  We use Eq.~\ref{eq:unitprod} to estimate the 3Q composite unitarity assuming no entangling errors between the constituent systems.  While it is challenging to measure the full 3Q unitarity directly, we can place an upper bound on it given the coherence limit and a lower bound given by $u_\Eop^{\rm p}$.   As $u_\Eop^{\rm p}$ converges to the coherence limit we can infer that negligible spectator entanglement is generated by the CR gate.  This is further evidenced by the fact that
the 1Q and 2Q unitarities converge to their respective coherence limits with the amplitude of target rotary pulsing.\\

Entanglement generated by the rotary echo can also be quantified by the unitary entanglement~\cite{Zanardi2001,Wang2002}. Suppose $U$ is a unitary acting on the bipartite space $\mathcal{H}_1\otimes \mathcal{H}_2$. $U$ can be naturally associated with a state $|\psi_U\rangle$ in $\mathcal{H}_1^{\otimes 2} \otimes \mathcal{H}_2^{\otimes 2}$ and the unitary entanglement of $U$, $E(U)$, is given by 
\begin{align}
E(U) &= 1-\text{tr}\left(\text{tr}_{3,4}( |\psi_U\rangle\langle \psi_U|)^2\right),
\end{align}
where $``\text{tr}_{3,4}"$ denotes the partial trace over subsystems 3,4 and $\text{tr}\left(\text{tr}_{3,4}( |\psi_U\rangle\langle \psi_U|)^2\right)$ is the purity of the reduced state. $E(U)$ can be computed directly from $U$ via~\cite{Wang2002},
\begin{align}
E(U)&=1-\frac{1}{d_1^2d_2^2} \text{tr}\left(\left(U^\dagger \right)^{\otimes 2} T_{1,3} U^{\otimes 2} T_{1,3}\right),
\end{align}
where $T_{1,3}$ is the permutation operator that swaps subsystems 1 and 3.  In Fig. ~\ref{figure:Fig3}, we calculate $E(U)$ as a function of rotary amplitude from the unitary generated by $H = \tilde{\nu}_{IYZ}\frac{IYZ}{2}+\tilde{\nu}_{IZZ}\frac{IZZ}{2}$, using the values of $\tilde{\nu}_{IYZ}$ and $\tilde{\nu}_{IZZ}$ measured in Fig. \ref{figure:Fig2} as a function of rotary amplitude.


\section{Quantum Volume}
While specialized gate sequences and multi-qubit RB are essential tools for gate calibration and characterization, to study the combined circuit improvement from both two-qubit errors and target-target spectator interactions we turn to Quantum Volume (QV)~\cite{Cross2019}. The QV metric is a single-parameter alternative to quantifying performance in terms of many individual parameters, such as gate fidelity, number of qubits, and qubit connectivity, and requires measurement of heavy output probability (HOP) greater than $\frac{2}{3}$.\\

To explore the effect of target rotary on QV, we execute QV circuits for two calibration test conditions - with and without target rotary. As we see a broad improvement in the spectator error with the addition of target rotary for a range of rotary amplitudes, in this experiment we calibrate the amplitude of the rotary pulse to produce a $2\pi$ rotation during each cross resonance pulse. \\

We report in Table \ref{table:QVtable} a comparison in HOP obtained for these two conditions. We find that HOP on subset A (see Appendix~\ref{sec:Volume} for details) successfully passes the $\frac{2}{3}$  HOP threshold with the addition of target rotary to achieve QV32 but does not when using the standard CR sequence. For more statistics, we executed a smaller sample of QV circuits across all five-qubit linear subsets on \emph{$ibmq\_johannesburg$} for the same two calibration conditions (Appendix~\ref{sec:Volume} for details). From this experiment, we measured an average increase in HOP of 0.013 with target rotary - demonstrating the benefit of this technique for our most comprehensive performance metric. 

\begin{table}[ht!]
\begin{center}
 \begin{tabular}{|p{0.1\textwidth}||p{0.11\textwidth}|p{0.11\textwidth}|p{0.11\textwidth}|}
 \hline
  & Rotary  & No Rotary  & $\Delta$ \\ 
 \hline
 HOP on A & $0.69\pm0.01 $ & $0.655\pm0.01 $ & $0.035\pm0.014 $ \\ 
 \hline
 Avg. HOP & $0.648\pm0.004$ & $0.635\pm0.004$ & $0.013\pm0.006$  \\
 \hline
\end{tabular}
\end{center}
\caption{Heavy output probability (HOP) of QV32 circuits for two calibration conditions - echoed CR with and without added target rotary - with difference in HOP ($\Delta$) presented in the last column. Details on subset A can be found in Appendix~\ref{sec:Volume}.}
\label{table:QVtable}
\end{table}
\section{Conclusion}
We identified and characterized higher order errors arising from static $ZZ$ and developed specialized tomography sequences (HEAT) to amplify and characterize these errors. We have shown the addition of target rotary pulses during the echoed CR gate can provide the simultaneous benefit of addressing single-qubit errors identified in the echoed CR Hamiltonian arising from static $ZZ$, reducing Hamiltonian error due to CR classical crosstalk, and suppressing unwanted entanglement with target spectator qubits due to static coupling, all without increasing the length of the echo sequence. In particular, a rotary tone can eliminate $\tilde{\nu}_{IYI}$ while being large enough to sufficiently suppress $\tilde{\nu}_{ZYI}$ and $\tilde{\nu}_{ZZI}$ as shown in Fig. \ref{figure:Fig1}(c), as well as $\tilde{\nu}_{IYZ}$ and $\tilde{\nu}_{IZZ}$ as shown in Fig. \ref{figure:Fig2}(c). Any residual $\tilde{\nu}_{IZI}$ and $\tilde{\nu}_{IYI}$ error can be compensated for with extra $Z$ rotations on the target when constructing a controlled-not. \\

We analyzed the impact of rotary via both gate error using randomized benchmarking and quantum volume circuit performance.  We found that the error mitigation offered by rotary pulsing brings our error metrics very close to the limit of coherence. Some coherent errors due to pulsing imperfections are not captured in this model, and we expect this topic to move closer to the forefront of device research as gates are pushed faster. 


\section{Acknowledgements}
We thank Oliver Dial, David McKay, William Shanks, and Matthias Steffen for helpful experimental discussions, as well as Seth Merkel, John Timmerwilke, and Lev Bishop for numerical and theoretical support. We thank Doug McClure, Michael Gordon, George Keefe, William Landers, Srikanth Srinivasan, and Cindy Wang for their work in creating and supporting the \emph{$ibmq\_johannesburg$}  system on which all data presented here was taken. This work was supported by the Army Research Office under contract W911NF-14-1-0124. Authors declare no competing interests.

\bibliographystyle{unsrt}
\bibliography{bibliography.bib}

\appendix
\setcounter{equation}{0}
\renewcommand{\thefigure}{A.\arabic{figure}}
\setcounter{figure}{0}
\renewcommand{\thetable}{A.\arabic{table}}
\setcounter{table}{0}


\section{Echo CR model}\label{sec:Echo}

The CR Hamiltonian can be modeled by
\begin{align}\label{eq:Ham}
H&= \sum_{j=0}^1 \left[\omega_j b_j^\dagger b_j + \frac{\delta_j}{2}b_j^\dagger b_j \left(b_j^\dagger b_j - \mathbbm{1}\right)\right] + J \left(b_0^\dagger b_1 + b_0 b_1^\dagger\right) \nonumber \\
&+ \Omega \cos(\omega_d t + \phi_C)\left(b_0^\dagger + b_0\right),
\end{align}
where we have set $\hbar=1$. The first transmon is designated the ``control" and the second is designated the ``target". The transmons are represented by Duffing oscillators $\omega_j b_j^\dagger b_j + \frac{\delta_j}{2}b_j^\dagger b_j \left(b_j^\dagger b_j - \mathbbm{1}\right)$ with frequency $\omega_j$ and non-linearity $\delta_j$. The coupling is a Jaynes-Cummings interaction $J \left(b_0^\dagger b_1 + b_0 b_1^\dagger\right)$ and the drive term is $\Omega \cos(\omega_d t + \phi)\left(b_0^\dagger + b_0\right)$ where $\Omega$, $\omega_d$, and $\phi$ are the drive amplitude, frequency, and phase respectively. \\

As outlined in~\cite{Magesan2020} an effective time-independent block-diagonal Hamiltonian can be obtained by the following procedure: diagonalize the free part of the Hamiltonian, rotate the drive term into the dressed basis, move into the frame rotating at the dressed target frequency, make the rotating-wave approximation (RWA), and finally block-diagonalize under the principle of least action. The block-diagonal model is valid provided the eigenvectors of the Hamiltonian prior to block-diagonalization do not have significant overlap connecting the different blocks~\cite{Magesan2020}. In particular, the frequencies of the transmons should be reasonably far from resonance-collisions.\\ 

The resulting block-diagonal Hamiltonian on the qubit-qubit subspace only has $I$ and $Z$ terms on the control qubit and takes the form
\begin{align}\label{eq:Hp}
H(\Omega) &= \nu_{IX} \frac{IX}{2} + \nu_{IZ} \frac{IZ}{2}  + \nu_{ZI} \frac{ZI}{2} \nonumber \\
&+ \nu_{ZX} \frac{ZX}{2}  + \nu_{ZZ} \frac{ZZ}{2},
\end{align}
where $\nu_j = \nu_j(\Omega)$. There are no $\nu_{IY}$ or $\nu_{ZY}$ terms as the rotation on the target qubit is purely in the X quadrature. Reversing the sign of the tone only reverses the sign of the non-diagonal Pauli coefficients since diagonal terms are even order in $\Omega$ while non-diagonal terms are odd order in $\Omega$,
\begin{align}\label{eq:Hm}
H(-\Omega) &= -\nu_{IX} \frac{IX}{2} + \nu_{IZ} \frac{IZ}{2}  + \nu_{ZI} \frac{ZI}{2} \nonumber \\
&- \nu_{ZX} \frac{ZX}{2}  + \nu_{ZZ} \frac{ZZ}{2}.
\end{align}

The Hamiltonian model of Eq.~\ref{eq:Hp} contains unwanted terms that create errors in implementing the ideal $ZX_{\frac{\pi}{2}}$ gate. One strategy for eliminating most of these errors is through the echoed $ZX_{\frac{\pi}{2}}$ gate,
\begin{align}
ZX_{\frac{\pi}{2}} &= XI \cdot ZX_{-\frac{\pi}{4}} \cdot XI  \cdot ZX_{\frac{\pi}{4}},
\end{align}
where a general rotation about $ZX$ by angle $\theta$, denoted $ZX_{\theta}$, is given by,
\begin{align}
ZX_{\theta} &= e^{-i \frac{\theta}{2} ZX}.
\end{align}
Assuming the time-independent Hamiltonian of Eq.~\ref{eq:Hp}, the $ZX_{\frac{\pi}{2}}$ gate is implemented via
\begin{align}\label{eq:Upulse}
U &= XI \cdot e^{-i H(-\Omega) t}\cdot XI \cdot e^{-i H(\Omega) t_g},
\end{align}
where $\Omega$ is the drive amplitude and $t$ is the gate time for $e^{-i H(\pm \Omega) t}$. Both of $e^{-i H(\pm \Omega) t}$ can be found analytically and $U$ in Eq.~\ref{eq:Upulse} can be modeled by,
\begin{align}\label{eq:Echo_unitary}
U &= A_{II} II + A_{IY} IY + A_{IZ} IZ + A_{ZX} ZX,
\end{align}
where
\begin{align}\label{eq:Echo_unitary_Pauli}
A_{II} &= \text{tr}(U II)/4 \nonumber \\ &=\frac{AB\cos(At/2)\cos(Bt/2)}{AB} \nonumber \\
&+  \frac{(\nu_{IX}^2-\nu_{IZ}^2)\sin(At/2)\sin(Bt/2)}{AB}\nonumber \\
&+  \frac{(-\nu_{ZX}^2+\nu_{ZZ}^2)\sin(At/2)\sin(Bt/2)}{AB},\nonumber
\end{align}
\begin{align}
A_{IY} &= \text{tr}(U IY)/4 \nonumber \\
&= -\frac{ 2i(\nu_{IX}\nu_{IZ}-\nu_{ZX}\nu_{ZZ})\sin(At/2)\sin(Bt/2)}{AB},\nonumber 
\end{align}
\begin{align}
A_{IZ} &= \text{tr}(U IZ)/4 \nonumber \\
&= -i\frac{(\nu_{IZ}-\nu_{ZZ})A\cos(At/2)\sin(Bt/2)}{AB}\nonumber \\
&-i\frac{B(\nu_{IZ}+\nu_{ZZ})\sin(At/2)\cos(Bt/2)}{AB},\nonumber
\end{align}
\begin{align}
A_{ZX} &= \text{tr}(U ZX)/4 \nonumber \\
&= i\frac{(\nu_{IX}-\nu_{ZX})A\cos(At/2)\sin(Bt/2)}{AB}\nonumber \\
&-i\frac{B(\nu_{IX}+\nu_{ZX})\sin(At/2)\cos(Bt/2)}{AB},\nonumber \\
\end{align}
and
\begin{align}
A&=\sqrt{(\nu_{IX}+\nu_{ZX})^2+(\nu_{IZ}+\nu_{ZZ})^2},\nonumber \\
B&=\sqrt{(\nu_{IX}-\nu_{ZX})^2+(\nu_{IZ}-\nu_{ZZ})^2}.
\end{align}

The Hamiltonian that generates the unitary is given by
\begin{align} \label{eq:HfromU}
H&=\frac{i\log(U)}{2t},
\end{align}
and the analytical expression for $H$ is given by
\begin{align}\label{eq:Ham_analytic}
H &= \tilde{\nu}_{II} \frac{II}{2}  + \tilde{\nu}_{IY}\frac{IY}{2} + \tilde{\nu}_{IZ} \frac{IZ}{2} + \tilde{\nu}_{ZX} \frac{ZX}{2} \nonumber \\
&= \frac{i}{2t}\left(B_{II} \frac{II}{2} + B_{IY}\frac{IY}{2} + B_{IZ} \frac{IZ}{2} + B_{ZX} \frac{ZX}{2}\right),
\end{align}
where 
\begin{align}
B_{II} &= \ln (A_{II}-M)+\ln(A_{II}+M),\nonumber \\
B_{IY} &= A_{IY}\left( \frac{-\ln(A_{II}-M) + \ln(A_{II}+M)}{M}\right), \nonumber \\
B_{IZ} &= A_{IZ}\left( \frac{-\ln(A_{II}-M) + \ln(A_{II}+M)}{M}\right), \nonumber \\
B_{ZX} &= A_{ZX}\left( \frac{-\ln(A_{II}-M) + \ln(A_{II}+M)}{M}\right), \nonumber \\
M&= \sqrt{A_{IY}^2+A_{IZ}^2+A_{ZX}^2}, \label{eq:AtoB}
\end{align}
and the $A_j$ are defined in Eq.~\ref{eq:Echo_unitary_Pauli}. Hence we see that the effective echoed Hamiltonian only has non-identity $IY$, $IZ$, and $ZX$ Pauli elements which depend on the underlying Hamiltonian parameters $\nu$ and time $t$. 

Due to the form of the problem we can define two unitary operators on the target qubit that depend on the state of the control,
\begin{align}
U_{|0\rangle} &= A_{II} I + A_{ZX} X + A_{IY} Y + A_{IZ} Z \nonumber \\
&= A_I^{|0\rangle} I + A_X^{|0\rangle} X + A_Y^{|0\rangle} Y + A_Z^{|0\rangle} Z, \nonumber \\
U_{|1\rangle} &= A_{II} I - A_{ZX} X + A_{IY} Y + A_{IZ} Z \nonumber \\
&= A_I^{|1\rangle} I + A_X^{|1\rangle} X + A_Y^{|1\rangle} Y + A_Z^{|1\rangle} Z.
\end{align}
Writing $U_{|j\rangle}$, $j \in \{0,1\}$ as elements of $SU(2)$ via
\begin{align}
U_{|j\rangle} &= e^{-i \frac{\theta_j}{2} \hat{n}_j \cdot \left(X,Y,Z\right)}, \nonumber
\end{align}
and noting for a given unitary $U$ the following holds,
\begin{align}
U_{1,1} &= \cos\left(\frac{\theta}{2}\right) - i \hat{n}_z \sin\left(\frac{\theta}{2}\right),\nonumber \\
U_{1,2} &= -(\hat{n}_y + i \hat{n}_x) \sin\left(\frac{\theta}{2}\right),\nonumber \\
U_{2,1} &= (\hat{n}_y - i \hat{n}_x) \sin\left(\frac{\theta}{2}\right),\nonumber \\
U_{2,2} &= \cos\left(\frac{\theta}{2}\right) + i \hat{n}_z \sin\left(\frac{\theta}{2}\right),\nonumber \\
\end{align}
we obtain for $j \in \{0,1\}$,
\begin{align}
\cos\left(\frac{\theta_j}{2}\right) &= A_I^{|j\rangle},\nonumber \\
-i\hat{n}_{j,x} \sin\left(\frac{\theta_j}{2}\right) &= A_X^{|j\rangle}, \nonumber \\
-i\hat{n}_{j,y} \sin\left(\frac{\theta_j}{2}\right) &= A_Y^{|j\rangle}, \nonumber \\
-i \hat{n}_{j,z} \sin\left(\frac{\theta_j}{2}\right) &= A_Z^{|j\rangle}.\nonumber
\end{align}

In the case of implementing a $ZX_{\frac{\pi}{2}}$ gate, since $\hat{n}$ is a unit vector and $\theta_0=\theta_1=\frac{\pi}{2}$, 
\begin{align}
M&=\frac{i}{\sqrt{2}},
\end{align}
and so 
\begin{align}
B_{II} &= \ln (A_{II}-M)+\ln(A_{II}+M) \nonumber \\
&= \ln \left(\frac{1-i}{\sqrt{2}}\right) + \ln \left(\frac{1+i}{\sqrt{2}}\right) = \frac{-i\pi}{4}+\frac{i\pi}{4} = 0,\nonumber
\end{align}
\begin{align}
B_{IY} &= A_{IY}\left( \frac{-\ln(A_{II}-M) + \ln(A_{II}+M)}{M}\right) \nonumber \\
&= A_{IY}\left(\frac{i\frac{\pi}{2}}{\frac{i}{\sqrt{2}}}\right) = A_{IY}\frac{\pi}{\sqrt{2}}, \nonumber
\end{align}
\begin{align}
B_{IZ} &= A_{IZ}\left( \frac{-\ln(A_{II}-M) + \ln(A_{II}+M)}{M}\right) \nonumber \\
&= A_{IZ}\left(\frac{i\frac{\pi}{2}}{\frac{i}{\sqrt{2}}}\right) = A_{IZ}\frac{\pi}{\sqrt{2}}, \nonumber 
\end{align}
\begin{align}
B_{ZX} &= A_{ZX}\left( \frac{-\ln(A_{II}-M) + \ln(A_{II}+M)}{M}\right) \nonumber \\
&= A_{ZX}\left(\frac{i\frac{\pi}{2}}{\frac{i}{\sqrt{2}}}\right) = A_{ZX}\frac{\pi}{\sqrt{2}}. \nonumber \\
\end{align}

\section{Echo CR model with rotary}\label{sec:Rotary_echo}

Including a rotary tone to Eq.~\ref{eq:Ham} gives
\begin{align}\label{eq:Ham_rot}
H&= \sum_{j=0}^1 \left[b_j^\dagger b_j + \frac{\delta_j}{2}b_j^\dagger b_j \left(b_j^\dagger b_j - \mathbbm{1}\right)\right] + J \left(b_0^\dagger b_1 + b_0 b_1^\dagger\right) \nonumber \\
&+ \Omega \cos(\omega_d t + \phi_C)\left(b_0^\dagger + b_0\right) \nonumber \\
&+ \Omega_R \cos(\omega_R t + \phi_R)\left(b_1^\dagger + b_1\right),
\end{align}
and the effective Hamiltonian takes the form
\begin{gather}\label{eq:Hp_rot}
H(\Omega,\Omega_R) = \nu_{IX} \frac{IX}{2} + \nu_{IY} \frac{IY}{2}  + \nu_{IZ} \frac{IZ}{2}  + \nu_{ZI} \frac{ZI}{2} \nonumber \\
 + \nu_{ZX} \frac{ZX}{2} + \nu_{ZY} \frac{ZY}{2} + \nu_{ZZ} \frac{ZZ}{2},
\end{gather}
where $\nu_j = \nu_j(\Omega,\Omega_R)$. As before, reversing the signs of the tones gives,
\begin{gather}\label{eq:Hm_rot}
H(-\Omega,-\Omega_R) = -\nu_{IX} \frac{IX}{2} - \nu_{IY} \frac{IY}{2}  + \nu_{IZ} \frac{IZ}{2}  + \nu_{ZI} \frac{ZI}{2} \nonumber \\
 - \nu_{ZX} \frac{ZX}{2} - \nu_{ZY} \frac{ZY}{2} + \nu_{ZZ} \frac{ZZ}{2}.
\end{gather}
In the limit of large $|\Omega_R|$, $\nu_{IX}$ grows unbounded as the rotary is a direct tone on the target. In addition, $\nu_{IZ}$ grows since off-resonant driving of higher levels of the target produces a phase shift on the computational subspace. The rotary tone has small impact on $\nu_{ZI}$ and $\nu_{ZX}$ so both remain effectively constant in $\Omega_R$. In addition, assuming $\phi_C=0$, both of $\nu_{IY}$ and $\nu_{ZY}$ are equal to 0. Lastly, $\nu_{ZZ}$ grows due to drive-induced ZZ from off-resonant driving of higher levels. 

As before, both $e^{-i H(\Omega) t}$ and $e^{-i H(-\Omega) t}$ can be found analytically with $U$ in Eq.~\ref{eq:Upulse} given by,
\begin{align}\label{eq:Echo_unitary_gen}
U &= A_{II} II + A_{IX} IX + A_{IY} IY + A_{IZ} IZ + A_{ZI} ZI \nonumber \\
&+ A_{ZX} ZX + A_{ZY} ZY + A_{ZZ} ZZ,
\end{align}
where
\begin{align}\label{eq:Echo_unitary_Pauli_rot}
A_{II} &= \text{tr}(U II)/4 \nonumber \\
&= \frac{AB\cos(At/2)\cos(Bt/2)}{AB}\nonumber \\
& + \frac{(\nu_{IX}^2 + \nu_{IY}^2-\nu_{IZ}^2)\sin(At/2)\sin(Bt/2)}{AB}\nonumber \\
& - \frac{(\nu_{ZX}^2+\nu_{ZY}^2-\nu_{ZZ}^2)\sin(At/2)\sin(Bt/2)}{AB},\nonumber
\end{align}
\begin{align}
A_{IX} &= \text{tr}(U IX)/4 \nonumber \\
&= \frac{ 2i(\nu_{IY}\nu_{IZ}-\nu_{ZY}\nu_{ZZ})\sin(At/2)\sin(Bt/2)}{AB},\nonumber
\end{align}
\begin{align}
A_{IY} &= \text{tr}(U IY)/4 \nonumber \\
&= -\frac{ 2i(\nu_{IX}\nu_{IZ}-\nu_{ZX}\nu_{ZZ})\sin(At/2)\sin(Bt/2)}{AB},\nonumber
\end{align}
\begin{align}
A_{IZ} &= \text{tr}(U IZ)/4 \nonumber \\
&= -i\frac{A(\nu_{IZ}-\nu_{ZZ})\cos(At/2)\sin(Bt/2)}{AB}\nonumber \\
& -i\frac{B(\nu_{IZ}+\nu_{ZZ})\sin(At/2)\cos(Bt/2)}{AB},\nonumber
\end{align}
\begin{align}
A_{ZI} &= \text{tr}(U ZI)/4 = 0, \nonumber
\end{align}
\begin{align}
A_{ZX} &= \text{tr}(U ZX)/4 \nonumber \\
& = i\frac{A(\nu_{IX}-\nu_{ZX})\cos(At/2)\sin(Bt/2)}{AB}\nonumber \\
& - i\frac{B(\nu_{IX}+\nu_{ZX})\sin(At/2)\cos(Bt/2)}{AB},\nonumber
\end{align}
\begin{align}
A_{ZY} &= \text{tr}(U ZY)/4 \nonumber \\
&= i\frac{A(\nu_{IY}-\nu_{ZY})\cos(At/2)\sin(Bt/2)}{AB}\nonumber \\
&= i\frac{- B(\nu_{IY}+\nu_{ZY})\sin(At/2)\cos(Bt/2)}{AB},\nonumber
\end{align}
\begin{align}
A_{ZZ} &= \text{tr}(U ZZ)/4 \nonumber \\
&= -\frac{ 2i(\nu_{IY}\nu_{ZX}-\nu_{IX}\nu_{ZY})\sin(At/2)\sin(Bt/2)}{AB},\nonumber \\
\end{align}
and
\begin{align}
A&=\sqrt{(\nu_{IX}+\nu_{ZX})^2+(\nu_{IY}+\nu_{ZY})^2+(\nu_{IZ}+\nu_{ZZ})^2},\nonumber \\
B&=\sqrt{(\nu_{IX}-\nu_{ZX})^2+(\nu_{IY}-\nu_{ZY})^2+(\nu_{IZ}-\nu_{ZZ})^2}.
\end{align}
The effective generating Hamiltonian is given by
\begin{align}\label{eq:Ham_analytic2}
H &= \tilde{\nu}_{II} \frac{II}{2} + \tilde{\nu}_{IX}\frac{IX}{2} + \tilde{\nu}_{IY}\frac{IY}{2} + \tilde{\nu}_{IZ} \frac{IZ}{2} \nonumber \\
&+ \tilde{\nu}_{ZI} \frac{ZI}{2}+ \tilde{\nu}_{ZX} \frac{ZX}{2} + \tilde{\nu}_{ZY} \frac{ZY}{2} + \tilde{\nu}_{ZZ} \frac{ZZ}{2}\nonumber \\
&=\frac{i}{2t}\Bigg(B_{II} \frac{II}{2} + B_{IX}\frac{IX}{2} + B_{IY}\frac{IY}{2} + B_{IZ} \frac{IZ}{2} \nonumber \\
&+ B_{ZI} \frac{ZI}{2}+ B_{ZX} \frac{ZX}{2} + B_{ZY} \frac{ZY}{2} + B_{ZZ} \frac{ZZ}{2}\Bigg),
\end{align}
with the $B$ coefficients given by,
\begin{align}\label{eq:Echo_Ham_Pauli_rot}
B_{II} &= \frac{1}{2}\left(\ln (A_{II}-M_2)+\ln (A_{II}+M_2)\right)\nonumber \\
& + \frac{1}{2}\left(\ln (A_{II}-M_1) + \ln (A_{II}+M_1)\right),\nonumber
\end{align}
\begin{align}
B_{IX} &= \frac{ (-A_{IX}+A_{ZX})M_1\ln(A_{II}-M_2) }{2M_1M_2}\nonumber \\
&+\frac{ (A_{IX}-A_{ZX})M_1\ln(A_{II}+M_2)}{2M_1M_2}\nonumber \\
&- \frac{(A_{IX}+A_{ZX})M_2(\ln(A_{II}-M_1)-\ln(A_{II}+M_1))}{2M_1M_2}, \nonumber
\end{align}
\begin{align}
B_{IY} &= \frac{ (-A_{IY}+A_{ZY})M_1\ln(A_{II}-M_2) }{2M_1M_2}\nonumber \\
&+\frac{ (A_{IY}-A_{ZY})M_1\ln(A_{II}+M_2)}{2M_1M_2}\nonumber \\
& - \frac{(A_{IY}+A_{ZY})M_2(\ln(A_{II}-M_1)-\ln(A_{II}+M_1))}{2M_1M_2}, \nonumber
\end{align}
\begin{align}
B_{IZ} &= \frac{ (-A_{IZ}+A_{ZZ})M_1\ln(A_{II}-M_2)}{2M_1M_2}\nonumber \\
&+ \frac{ (A_{IZ}-A_{ZZ})M_1\ln(A_{II}+M_2)}{2M_1M_2}\nonumber \\
& - \frac{(A_{IZ}+A_{ZZ})M_2(\ln(A_{II}-M_1)-\ln(A_{II}+M_1))}{2M_1M_2}, \nonumber
\end{align}
\begin{align}
B_{ZI} & = 0,\nonumber \\
\end{align}
\begin{align}
B_{ZX} &= \frac{ (A_{IX}-A_{ZX})M_1\ln(A_{II}-M_2) }{2M_1M_2}\nonumber \\
&+ \frac{ (-A_{IX}+A_{ZX})M_1\ln(A_{II}+M_2)}{2M_1M_2}\nonumber \\
& - \frac{(A_{IX}+A_{ZX})M_2(\ln(A_{II}-M_1)-\ln(A_{II}+M_1))}{2M_1M_2}, \nonumber
\end{align}
\begin{align}
B_{ZY} &= \frac{ (A_{IY}-A_{ZY})M_1\ln(A_{II}-M_2) }{2M_1M_2}\nonumber \\
&+ \frac{ (-A_{IY}+A_{ZY})M_1\ln(A_{II}+M_2)}{2M_1M_2}\nonumber \\
& - \frac{(A_{IY}+A_{ZY})M_2(\ln(A_{II}-M_1)-\ln(A_{II}+M_1))}{2M_1M_2}, \nonumber
\end{align}
\begin{align}
B_{ZZ} &= \frac{ (A_{IZ}-A_{ZZ})M_1\ln(A_{II}-M_2)}{2M_1M_2}\nonumber \\
&+ \frac{ (-A_{IZ}+A_{ZZ})M_1\ln(A_{II}+M_2)}{2M_1M_2}\nonumber \\
& - \frac{(A_{IZ}+A_{ZZ})M_2(\ln(A_{II}-M_1)-\ln(A_{II}+M_1))}{2M_1M_2}, 
\end{align}
where
\begin{align}
M_1&= \sqrt{(A_{IX}+A_{ZX})^2+(A_{IY}+A_{ZY})^2+(A_{IZ}+A_{ZZ})^2},\nonumber \\
M_2&= \sqrt{(A_{IX}-A_{ZX})^2+(A_{IY}-A_{ZY})^2+(A_{IZ}-A_{ZZ})^2},
\end{align}
and the $A_j$ are defined in Eq.~\ref{eq:Echo_unitary_Pauli_rot}.

Again note that in the case of implementing a $ZX_{\frac{\pi}{2}}$ gate (so that $\theta=\frac{\pi}{2}$),
\begin{align}
M_1&=M_2 = \frac{i}{\sqrt{2}}, \nonumber \\
B_{II}&=0,
\end{align}
and the non-identity $B$ and $A$ coefficients are related via
\begin{align}\label{eq:BtoA_rot}
B_{ij}&= \frac{\pi}{\sqrt{2}}A_{ij}.
\end{align}

\subsection{Eliminating $A_{IY}$}

One strategy for reducing the gate error is to try and eliminate $A_{IY}$ so that the rotary echo error is an $IZ$ rotation that can be corrected via a frame change. We see that $A_{IY}=0$ if one of the following is satisfied for neither of $A$ or $B$ equal to 0:
\begin{enumerate}
\item $\nu_{IX}\nu_{IZ}-\nu_{ZX}\nu_{ZZ} = 0$,
\item $A = \frac{2n\pi}{t}$, $n > 0$,
\item $B = \frac{2n\pi}{t}$, $n > 0$.
\end{enumerate}
We restrict $n > 0$ since $A$ and $B$ are non-negative and not equal to 0 by assumption. Note also that 
\begin{align}
B &= 0 \Rightarrow \nu_{IX}\nu_{IZ}-\nu_{ZX}\nu_{ZZ} = 0.
\end{align}
We define three classes of solutions,
\begin{align}
\chi_0 &:=\nu_{IX}\nu_{IZ}-\nu_{ZX}\nu_{ZZ}, \nonumber \\
\chi_{1,n} &:=A-\frac{2n\pi}{t}, \nonumber \\
\chi_{2,n} &:=B-\frac{2n\pi}{t},
\end{align}
and $A_{IY}=B_{IY}=0$ if and only if $\chi_0=0$, $\chi_{1,n}=0$, or $\chi_{2,n}=0$ for some $n > 0$.

\section{Echo CR model with crosstalk and rotary}\label{sec:Echo_ct_rotary}

Suppose there are both CR and rotary tones and classical crosstalk from the CR tone to the target. We assume crosstalk from the rotary tone back to the control transmon is negligible compared to crosstalk from the CR tone. The Hamiltonian for the system is given by
\begin{align}
H&= \sum_{j=0}^1 \left[b_j^\dagger b_j + \frac{\delta_j}{2}b_j^\dagger b_j \left(b_j^\dagger b_j - \mathbbm{1}\right)\right] + J \left(b_0^\dagger b_1 + b_0 b_1^\dagger\right)\nonumber \\
& + \Omega \cos(\omega_d t + \phi_C)\left(b_0^\dagger + b_0\right)\nonumber \\
& + \Omega_T \cos(\omega_d t + (\phi_C-\phi_T))\left(b_1^\dagger + b_1\right)\nonumber \\
& + \Omega_R \cos(\omega_d t + \phi_R)\left(b_1^\dagger + b_1\right),
\end{align}
where $\Omega \cos(\omega_d t + \phi_C)\left(b_0^\dagger + b_0\right)$ is the CR tone, $\Omega_T \cos(\omega_d t + (\phi_C-\phi_T))\left(b_1^\dagger + b_1\right)$ is the classical crosstalk, and $\Omega_R \cos(\omega_d t + \phi_R)\left(b_1^\dagger + b_1\right)$ is the rotary tone. Here $\phi_T$ is a constant that represents the phase accumulation due to the path length of the crosstalk signal seen by the target. We can re-write this Hamiltonian in the form of Eq.~\ref{eq:Ham_rot} by summing the crosstalk and rotary signals into a single cosine,
\begin{align}
H&= \sum_{j=0}^1 \left[b_j^\dagger b_j + \frac{\delta_j}{2}b_j^\dagger b_j \left(b_j^\dagger b_j - \mathbbm{1}\right)\right] + J \left(b_0^\dagger b_1 + b_0 b_1^\dagger\right)\nonumber \\
& + \Omega \cos(\omega_d t)\left(b_0^\dagger + b_0\right) + \tilde{\Omega} \cos(\omega_d t + \tilde{\phi})\left(b_1^\dagger + b_1\right),
\end{align}
where
\begin{align}
\tilde{\Omega}^2 &= \Omega_T^2 + \Omega_R^2 + 2\Omega_T\Omega_R \cos((-\phi_T)-\phi_R) \nonumber \\
&=\Omega_T^2 + \Omega_R^2 + 2\Omega_T\Omega_R \cos(\phi_T+\phi_R),\nonumber \\
\tilde{\phi} &= \arctan\left(\frac{\Omega_T\sin(-\phi_T) + \Omega_R\sin(\phi_R)}{\Omega_T\cos(-\phi_T) + \Omega_R\cos(\phi_R)}\right) \nonumber \\
&= \arctan\left(\frac{-\Omega_T\sin(\phi_T) + \Omega_R\sin(\phi_R)}{\Omega_T\cos(\phi_T) + \Omega_R\cos(\phi_R)}\right). \label{eq:Omega_phi_T}
\end{align}
Note that even if $\phi_R=0$ the amplitude $\Omega_R$ can still affect the resultant phase $\tilde{\phi}$. If the effective Hamiltonian has the form
\begin{gather}\label{eq:Hp_ct_rot}
H(\Omega,\Omega_T,\Omega_R) = H(\Omega,\tilde{\Omega}) = \nu_{IX} \frac{IX}{2} + \nu_{IY} \frac{IY}{2}  + \nu_{IZ} \frac{IZ}{2}\nonumber \\
  + \nu_{ZI} \frac{ZI}{2}  + \nu_{ZX} \frac{ZX}{2} + \nu_{ZY} \frac{ZY}{2} + \nu_{ZZ} \frac{ZZ}{2},
\end{gather}
where $\nu_j = \nu_j(\Omega,\tilde{\Omega}) = \nu_j(\Omega,\Omega_T,\Omega_R)$ then since changing the sign of $\Omega_T$ and $\Omega_R$ changes the sign of $\tilde{\Omega}$, from the solution of Eq.~\ref{eq:Omega_phi_T},
\begin{gather}\label{eq:Hm_ct_rot}
H(-\Omega,-\Omega_T,-\Omega_R) = H(-\Omega,-\tilde{\Omega}) = -\nu_{IX} \frac{IX}{2} - \nu_{IY} \frac{IY}{2} \nonumber \\
 + \nu_{IZ} \frac{IZ}{2} + \nu_{ZI} \frac{ZI}{2}  - \nu_{ZX} \frac{ZX}{2} - \nu_{ZY} \frac{ZY}{2} + \nu_{ZZ} \frac{ZZ}{2}.
\end{gather}

\subsection{Behavior of $\nu$ and $A$ coefficients in large rotary amplitude limit}\label{sec:nu_ct_rotary}

We look at the realistic limit where the rotary amplitude is much larger than the crosstalk amplitude, $|\Omega_R| \gg |\Omega_T|$, and set $\phi_C=\phi_R=0$. As in the case with no crosstalk, $\nu_{IX}$ grows unbounded, $\nu_{IZ}$ grows from phase accumulated due to higher levels present, $\nu_{ZI}$ and $\nu_{ZX}$ remain effectively constant, $\nu_{ZY}$ is equal to 0 since $\phi_C=0$, and $\nu_{ZZ}$ grows due to drive-induced $ZZ$. What remains is understanding the behavior of $\nu_{IY}$. 

Since $\phi_C=\phi_R=0$,
\begin{align} 
\tilde{\Omega}^2 &= \Omega_T^2 + \Omega_R^2 + 2\Omega_T\Omega_R \cos(\phi_T),\nonumber \\
\tilde{\phi} &= \arctan\left(\frac{-\Omega_T\sin(\phi_T)}{\Omega_T\cos(\phi_T) + \Omega_R}\right).
\end{align}
As $|\Omega_R| \gg |\Omega_T|$ we rewrite the above as
\begin{align} 
\tilde{\Omega}^2 &= \Omega_R^2\left(1+ \frac{\Omega_T^2}{\Omega_R^2} + \frac{2\Omega_T}{\Omega_R} \cos(\phi_T)\right)\nonumber \\
&= \Omega_R^2\left(1+ \epsilon^2 + 2\epsilon \cos(\phi_T)\right),\nonumber \\
\tilde{\phi} &= \arctan\left(\frac{-\Omega_T\sin(\phi_T)}{\Omega_R\left(1+\frac{\Omega_T}{\Omega_R}\cos(\phi_T)\right)}\right) \nonumber \\
&= -\arctan\left(\frac{\epsilon \sin(\phi_T)}{1+\epsilon\cos(\phi_T)}\right),
\end{align}
where $\epsilon = \frac{\Omega_T}{\Omega_R}$. For the amplitude we have
\begin{align} 
\tilde{\Omega} &= \pm |\Omega_R|\sqrt{\left(1+ \epsilon^2 + 2\epsilon \cos(\phi_T)\right)},
\end{align}
where if $\Omega_R \gg 0$ we take the positive solution and if $\Omega_R \ll 0$ we take the negative solution. As $\phi_C=0$, we can approximate $\nu_{IY}$ by $-\tilde{\Omega}\sin(\tilde{\phi})$. Hence to first order in $\epsilon$ if $\Omega_R \gg 0$,
\begin{align}
\nu_{IY} \approx -\epsilon |\Omega_R| \sin(\phi_T) = -\Omega_T\sin(\phi_T),
\end{align}
while if $\Omega_R \ll 0$,
\begin{align}
\nu_{IY} \approx \epsilon |\Omega_R| \sin(\phi_T) = -\Omega_T\sin(\phi_T).
\end{align}
In the case of crosstalk with no rotary, $\left(\frac{IY}{2}\right)_\text{coeff} \approx -\Omega_T \sin(\phi_T)$. Hence we see that the $\nu_{IY}$ coefficient is effectively unchanged by the addition of the rotary tone. 

Summarizing, assuming $\phi_C=\phi_R=0$ we have that in the limit of $|\Omega_R|$ growing large the $\nu$ parameters are given by:
\begin{enumerate}
\item $\nu_{IX}$ grows large in magnitude (driving target on-resonance),
\item $\nu_{IY}$ remains effectively constant (shown above),
\item $\nu_{IZ}$ grows large in magnitude (phase shift from off-resonant driving of higher levels),
\item $\nu_{ZI}$ remains effectively constant,
\item $\nu_{ZX}$ remains effectively constant,
\item $\nu_{ZY}$ remains effectively constant and equal to 0,
\item $\nu_{ZZ}$ grows large in magnitude (phase shift from off-resonant driving of higher levels),
\end{enumerate}
and in this limit $\nu_{IX}$ dominates all other terms in magnitude. 

Next, let us look at the behavior of the $A$ coefficients in the large $\Omega_R$ limit. From Eq.~\ref{eq:Echo_unitary_Pauli_rot} (still assuming $\phi_C=\phi_R=0$) we see that the components that do not have $\nu_{IX}$, $\nu_{IZ}$, or $\nu_{ZZ}$ in the numerator will damp to 0 in the large $|\Omega_R|$ limit. These include $A_{ZY}$ and $A_{ZZ}$. We expect $A_{IX}$ to be small since $\nu_{ZY}=0$ and the product $\nu_{IY}\nu_{IZ}$ is small (but will grow with $\nu_{IZ}$). Next, $A_{IY}$ and $A_{IZ}$ both have terms growing large in the numerator and denominator so we don't expect them to damp out from the application of the rotary tone. Lastly we expect $A_{ZX}$ to be relatively independent of the rotary amplitude and be the main coefficient in the large rotary limit.

\section{Hamiltonian error amplifying tomography (HEAT): Rotary echo}\label{sec:HEAT_re}

For rotary echo we have that the unitary $U$ describing the evolution is given by,
\begin{align}
U&=A_{II}II + A_{IX}IX + A_{IY}IY + A_{IZ}IZ \nonumber \\
&+ A_{ZX} ZX + A_{ZY} ZY + A_{ZZ} ZZ,
\end{align}
where we include $ZY$ and $ZZ$ terms in the case of crosstalk or phase misalignment. If the control is initially in $|0\rangle$ then the evolution of the target qubit is described by
\begin{align}
U_{|0\rangle} &= A_I^{|0\rangle}I + A_Y^{|0\rangle}Y + A_Z^{|0\rangle}Z + A_X^{|0\rangle} X \nonumber \\
&= A_{II}I + (A_{IY} + A_{ZY})Y + (A_{IZ}+A_{ZZ})Z \nonumber \\
&+ (A_{IX}+A_{ZX}) X,
\end{align}
while if the control is initially in $|1\rangle$ then
\begin{align}
U_{|1\rangle} &= A_I^{|1\rangle}I + A_Y^{|1\rangle}Y + A_Z^{|1\rangle}Z + A_X^{|1\rangle} X \nonumber \\
&= A_{II}I + (A_{IY} - A_{ZY})Y + (A_{IZ} - A_{ZZ})Z \nonumber \\
&+ (A_{IX}- A_{ZX})X.
\end{align}

An element of $SU(2)$ takes the form
\begin{align}
e^{-i \frac{\theta}{2} \hat{n}\cdot (X,Y,Z)} &= \cos\left(\frac{\theta}{2}\right) I - i\sin\left(\frac{\theta}{2}\right) \hat{n}\cdot (X,Y,Z),
\end{align}
and a given unitary $U \in SU(2)$ can be written in this form via
\begin{align}
U_{1,1} &= \cos\left(\frac{\theta}{2}\right) - i \hat{n}_z \sin\left(\frac{\theta}{2}\right),\nonumber \\
U_{1,2} &= -(\hat{n}_y + i \hat{n}_x) \sin\left(\frac{\theta}{2}\right),\nonumber \\
U_{2,1} &= (\hat{n}_y - i \hat{n}_x) \sin\left(\frac{\theta}{2}\right),\nonumber \\
U_{2,2} &= \cos\left(\frac{\theta}{2}\right) + i \hat{n}_z \sin\left(\frac{\theta}{2}\right).\nonumber \\
\end{align}
Hence for $j \in \{0,1\}$,
\begin{align}
U_{|j\rangle} &= e^{-i \frac{\theta_j}{2} \hat{n}_j \cdot (X,Y,Z)},
\end{align}
where
\begin{align}
\cos\left(\frac{\theta_j}{2}\right) &= A_I^{|j\rangle},\nonumber \\
-i\hat{n}_{j,x} \sin\left(\frac{\theta_j}{2}\right) &= A_X^{|j\rangle}, \nonumber \\
-i\hat{n}_{j,y} \sin\left(\frac{\theta_j}{2}\right) &= A_Y^{|j\rangle}, \nonumber \\
-i \hat{n}_{j,z} \sin\left(\frac{\theta_j}{2}\right) &= A_Z^{|j\rangle}.\nonumber
\end{align}

\begin{figure}[h!]
\centering
\includegraphics[width=0.49\textwidth]{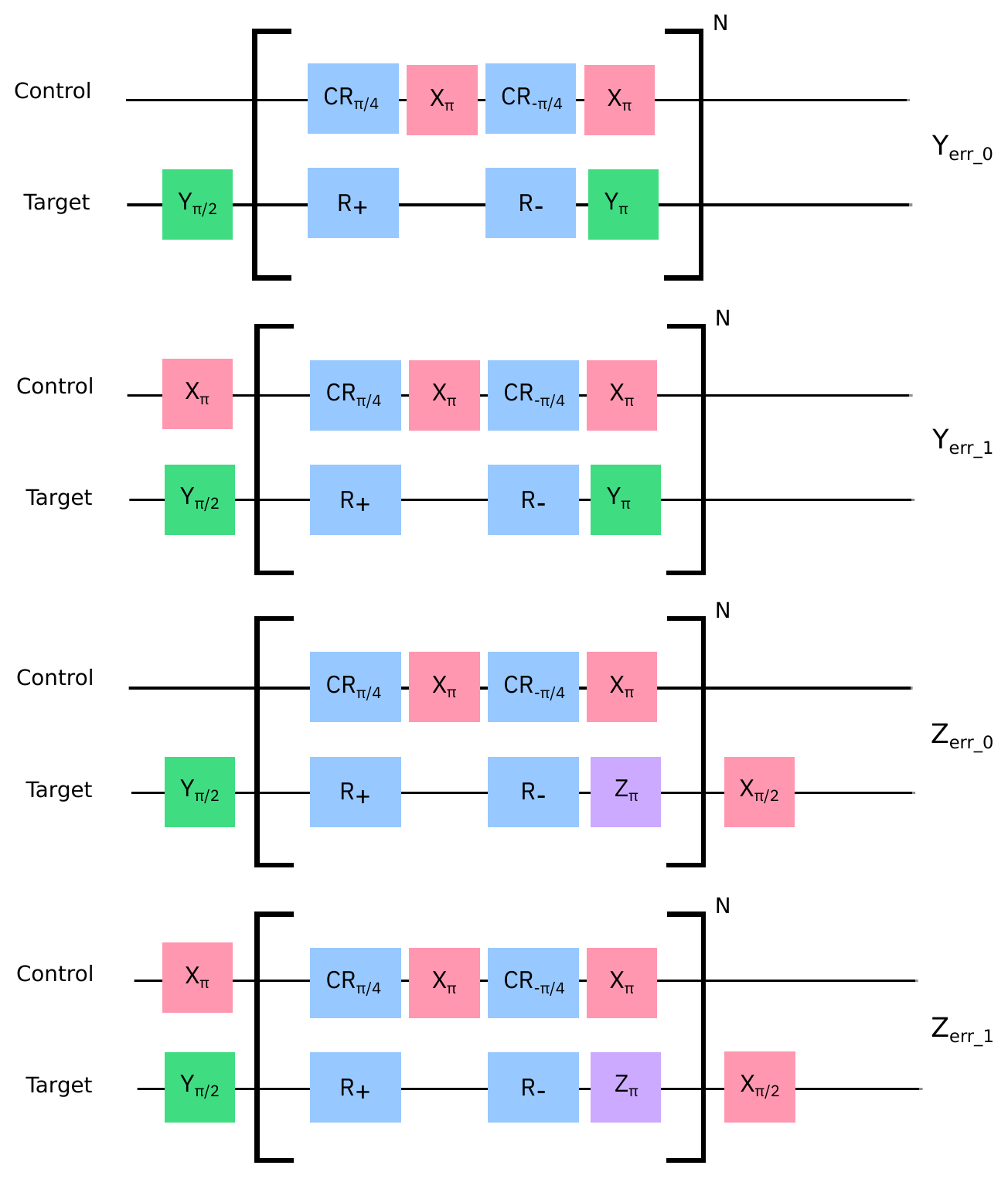}
\caption{\label{fig:YpZpSeq}  HEAT sequences for reconstructing error terms.}
\end{figure}

The set of HEAT sequences are given by Fig.~\ref{fig:YpZpSeq}.
To first order in $\hat{n}_{0,y}$ and $\hat{n}_{0,z}$ and for even $N$ (similarly for $\hat{n}_{1,y}$ and $\hat{n}_{1,z}$ since the sequences do not distinguish between preparation of the control in $|0\rangle$ or $|1\rangle$) the measurement expectation values of the output state $\rho_N$ from the $N$ repetitions is given by
\begin{align}
Y_{\text{err}_{0,N}} &= \text{tr}(\rho_N Z) \approx  -\frac{N\hat{n}_{0,y}\sin(\theta_0)}{|\hat{n}_{0,x}|}, \nonumber \\
Z_{\text{err}_{0,N}} &= \text{tr}(\rho_N Y) \approx  \frac{N\hat{n}_{0,z}\sin(\theta_0)}{|\hat{n}_{0,x}|},
\end{align}
while for odd $N$ there is mixing between $\hat{n}_{0,Y}$ and $\hat{n}_{0,Z}$. The above holds for $\hat{n}_{1,X}$, $\hat{n}_{1,Y}$, $\hat{n}_{1,Z}$, and $\theta_1$ so that for even $N$,
\begin{align}
Y_{\text{err}_{1,N}} &= \text{tr}(\rho_N Z) \approx  -\frac{N\hat{n}_{1,y}\sin(\theta_1)}{|\hat{n}_{1,x}|}, \nonumber \\
Z_{\text{err}_{1,N}} &= \text{tr}(\rho_N Y) \approx  \frac{N\hat{n}_{1,z}\sin(\theta_1)}{|\hat{n}_{1,x}|}.
\end{align}
$A_{II}$, $A_{IX}$, $A_{IY}$, $A_{IZ}$, $A_{ZX}$, $A_{ZY}$, and $A_{ZZ}$ can be reconstructed from the above equations. First, note that for each $j=0,1$,
\begin{align}
\sum_{a=1}^3 \hat{n}_{j,a}^2&=1,
\end{align}
so that 
\begin{align}
|\hat{n}_{j,x}|^2\left(1+\left(\frac{Y_{\text{err}_{j,N}}}{N\sin(\theta_j)}\right)^2+ \left(\frac{Z_{\text{err}_{j,N}}}{N\sin(\theta_j)}\right)^2\right)&=1.
\end{align}
Hence for $j=0,1$,
\begin{align}
\hat{n}_{j,x} &= \pm \sqrt{\frac{1}{\left(1+\left(\frac{Y_{\text{err}_{j,N}}}{N\sin(\theta_j)}\right)^2+ \left(\frac{Z_{\text{err}_{j,N}}}{N\sin(\theta_j)}\right)^2\right)}},
\end{align}
and for the $y$ and $z$ components,
\begin{align}
\hat{n}_{0,y} &\approx \left(-\frac{Y_{\text{err}_{0,N}}}{N\sin(\theta_0)}\right)|\hat{n}_{0,x}|, \nonumber \\
\hat{n}_{1,y} &\approx \left(-\frac{Y_{\text{err}_{1,N}}}{N\sin(\theta_1)}\right)|\hat{n}_{1,x}|, \nonumber \\
\hat{n}_{0,z} &\approx \left(\frac{Z_{\text{err}_{0,N}}}{N\sin(\theta_0)}\right)|\hat{n}_{0,x}|, \nonumber \\
\hat{n}_{1,z} &\approx \left(\frac{Z_{\text{err}_{1,N}}}{N\sin(\theta_1)}\right)|\hat{n}_{1,x}|.
\end{align}
Using these values one can reconstruct the $A$ parameters. 

As an example, in the case of implementing a $ZX_{\frac{\pi}{2}}$ gate,
\begin{align}
\theta_0&=\theta_1 \approx \frac{\pi}{2}, \nonumber \\
\hat{n}_{0,x} &\approx \sqrt{\frac{1}{\left(1+\left(\frac{Y_{\text{err}_{0,N}}}{N\sin(\theta_0)}\right)^2+ \left(\frac{Z_{\text{err}_{0,N}}}{N\sin(\theta_0)}\right)^2\right)}} \nonumber \\
& \approx \sqrt{\frac{1}{\left(1+\left(\frac{Y_{\text{err}_{0,N}}}{N}\right)^2+ \left(\frac{Z_{\text{err}_{0,N}}}{N}\right)^2\right)}} \nonumber \\
\hat{n}_{1,x} &\approx -\sqrt{\frac{1}{\left(1+\left(\frac{Y_{\text{err}_{1,N}}}{N\sin(\theta_1)}\right)^2+ \left(\frac{Z_{\text{err}_{1,N}}}{N\sin(\theta_1)}\right)^2\right)}} \nonumber \\
&\approx -\sqrt{\frac{1}{\left(1+\left(\frac{Y_{\text{err}_{1,N}}}{N}\right)^2+ \left(\frac{Z_{\text{err}_{1,N}}}{N}\right)^2\right)}},
\end{align}
\begin{align}
A_{II} &=  \cos\left(\frac{\theta_0}{2}\right) = \cos\left(\frac{\theta_1}{2}\right) \approx \frac{1}{\sqrt{2}}, \nonumber \\
A_{IX} &=  \frac{A_X^{|0\rangle}+A_X^{|1\rangle}}{2} \approx \frac{-i}{2\sqrt{2}}(\hat{n}_{0,x} + \hat{n}_{1,x}), \nonumber \\
A_{IY} &=  \frac{A_Y^{|0\rangle}+A_Y^{|1\rangle}}{2} \approx \frac{-i}{2\sqrt{2}}(\hat{n}_{0,y} + \hat{n}_{1,y}), \nonumber \\
A_{IZ} &=  \frac{A_Z^{|0\rangle}+A_Z^{|1\rangle}}{2} \approx \frac{-i}{2\sqrt{2}}(\hat{n}_{0,z} + \hat{n}_{1,z}), \nonumber \\
A_{ZX} &=  \frac{A_X^{|0\rangle}-A_X^{|1\rangle}}{2} \approx \frac{-i}{2\sqrt{2}}(\hat{n}_{0,x} - \hat{n}_{1,x}), \nonumber \\
A_{ZY} &=  \frac{A_Y^{|0\rangle}-A_Y^{|1\rangle}}{2} \approx \frac{-i}{2\sqrt{2}}(\hat{n}_{0,y} - \hat{n}_{1,y}), \nonumber \\
A_{ZZ} &=  \frac{A_Z^{|0\rangle}-A_Z^{|1\rangle}}{2} \approx \frac{-i}{2\sqrt{2}}(\hat{n}_{0,z} - \hat{n}_{1,z}). \nonumber \\
\end{align}
Assuming $A_{ZI}=0$ the Hamiltonian $B$ coefficients can now be reconstructed from Eq.~\ref{eq:Echo_Ham_Pauli_rot}.

\section{Hamiltonian error amplifying tomography: Spectator-Target case}\label{sec:HEAT_st}

Assume $U$ has the form
\begin{align}
U&=  A_{II}II + A_{IY}IY + A_{IZ}IZ + A_{ZY}ZY + A_{ZZ}ZZ.
\end{align}
If the spectator is initially in $|0\rangle$ then the evolution on the target qubit is described by
\begin{align}
U_{|0\rangle} &= A_I^{|0\rangle}I + A_Y^{|0\rangle}Y + A_Z^{|0\rangle}Z\nonumber \\
&= A_{II}I + (A_{IY}+A_{ZY})Y + (A_{IZ}+A_{ZZ})Z,
\end{align}
while if the control is initially in $|1\rangle$ then
\begin{align}
U_{|1\rangle} &= A_I^{|1\rangle}I + A_Y^{|1\rangle}Y + A_Z^{|1\rangle}Z\nonumber \\
&= A_{II}I + (A_{IY}-A_{ZY})Y + (A_{IZ}-A_{ZZ})Z.
\end{align}
As before, $U_{|0\rangle}$ can be written as
\begin{align}
U_{|0\rangle} &= e^{-i \frac{\theta_0}{2} \hat{n_0}\cdot (X,Y,Z)} \nonumber \\
&= \cos\left(\frac{\theta_0}{2}\right) I - i\sin\left(\frac{\theta_0}{2}\right) \hat{n_0}\cdot (X,Y,Z),
\end{align}
and noting $\hat{n}_{0,x}=0$,
\begin{align}\label{eq:ntoA_0}
\cos\left(\frac{\theta_0}{2}\right) &= A_I^{|0\rangle} = A_{II},\nonumber \\
\hat{n}_{0,y} \sin\left(\frac{\theta_0}{2}\right) &= iA_Y^{|0\rangle} = i(A_{IY}+A_{ZY}), \nonumber \\
\hat{n}_{0,z} \sin\left(\frac{\theta_0}{2}\right) &= iA_Z^{|0\rangle} = i(A_{IZ}+A_{ZZ}).\nonumber \\
\end{align}
Similarly for $U_{|1\rangle}$,
\begin{align}
U_{|1\rangle} &= e^{-i \frac{\theta_1}{2} \hat{n}_1 \cdot (X,Y,Z)}\nonumber \\
&= \cos\left(\frac{\theta_1}{2}\right) I - i\sin\left(\frac{\theta_1}{2}\right) \hat{n_1}\cdot (X,Y,Z),
\end{align}
with
\begin{align}\label{eq:ntoA_1}
\cos\left(\frac{\theta_1}{2}\right) &= A_I^{|1\rangle} = A_{II},\nonumber \\
\hat{n}_{1,y} \sin\left(\frac{\theta_1}{2}\right) &= iA_Y^{|1\rangle} = i(A_{IY}-A_{ZY}), \nonumber \\
\hat{n}_{1,z} \sin\left(\frac{\theta_1}{2}\right) &= iA_Z^{|1\rangle} = i(A_{IZ}-A_{ZZ}).
\end{align}

In the small $\theta$ limit we can write,
\begin{align}
\hat{n}_{0,y} \sin\left(\frac{\theta_0}{2}\right) &\approx \hat{n}_{0,y} \frac{\theta_0}{2}, \nonumber \\
\hat{n}_{0,z} \sin\left(\frac{\theta_0}{2}\right) &\approx \hat{n}_{0,z} \frac{\theta_0}{2}, \nonumber \\
\hat{n}_{1,y} \sin\left(\frac{\theta_1}{2}\right) &\approx \hat{n}_{1,y} \frac{\theta_1}{2}, \nonumber \\
\hat{n}_{1,z} \sin\left(\frac{\theta_1}{2}\right) &\approx \hat{n}_{1,z} \frac{\theta_1}{2}.
\end{align}
Using the sequences in Fig.~\ref{fig:YpZpSeq} we find in this case that for even $N$ and small $\theta_0$,
\begin{align}\label{eq:Seq0err}
Y_{\text{err}_{0,N}} &= \text{tr}(\rho_N Z) \approx - N\hat{n}_{0,y}\sin(\theta_0) \approx -N\hat{n}_{0,y} \theta_0, \nonumber \\
Z_{\text{err}_{0,N}} &= \text{tr}(\rho_N Y) \approx  N\hat{n}_{0,z}\sin(\theta_0) \approx N\hat{n}_{0,z} \theta_0,
\end{align}
while for odd $N$ there is mixing between $\hat{n}_{0,y}$ and $\hat{n}_{0,z}$. Analogous results hold for $\hat{n}_{1,Y}$ and $\hat{n}_{1,Z}$: for even $N$ and small $\theta_1$,
\begin{align}\label{eq:Seq1err}
Y_{\text{err}_{1,N}} &= \text{tr}(\rho_N Z) \approx - N\hat{n}_{1,y}\sin(\theta_1) \approx -N\hat{n}_{1,y} \theta_1, \nonumber \\
Z_{\text{err}_{1,N}} &= \text{tr}(\rho_N Y) \approx  N\hat{n}_{1,z}\sin(\theta_1) \approx N\hat{n}_{1,z} \theta_1.
\end{align}
One can therefore estimate the $A$ terms in this limit by using Eq.'s~\ref{eq:ntoA_0}, ~\ref{eq:ntoA_1},~\ref{eq:Seq0err}, and~\ref{eq:Seq1err},
\begin{align}
A_{IY} + A_{ZY} &= -i \hat{n}_{0,y} \sin\left(\frac{\theta_0}{2}\right)\nonumber \\
&\approx \frac{-i \hat{n}_{0,y}\theta_0}{2} \approx \frac{i\text{tr}(\rho_N Z)}{2N}, \nonumber \\
A_{IZ} + A_{ZZ} &= -i \hat{n}_{0,z} \sin\left(\frac{\theta_0}{2}\right) \nonumber \\
&\approx \frac{-i \hat{n}_{0,z}\theta_0}{2} \approx \frac{-i\text{tr}(\rho_N Y)}{2N}, \nonumber \\
A_{IY} - A_{ZY} &= -i \hat{n}_{1,y} \sin\left(\frac{\theta_1}{2}\right) \nonumber \\
&\approx \frac{-i \hat{n}_{1,y}\theta_1}{2} \approx \frac{i\text{tr}(\rho_N Z)}{2N}, \nonumber \\
A_{IZ} - A_{ZZ} &= -i \hat{n}_{1,z} \sin\left(\frac{\theta_1}{2}\right) \nonumber \\
&\approx \frac{-i \hat{n}_{1,z}\theta_1}{2} \approx \frac{-i\text{tr}(\rho_N Y)}{2N}. \nonumber \\
\end{align}
$A_{II}$ can be computed by noting that
\begin{align}
\frac{\theta_0}{2} \approx \pm i \sqrt{(A_{IY}+A_{ZY})^2+(A_{IZ}+A_{ZZ})^2}, \nonumber \\
\frac{\theta_1}{2} \approx \pm i \sqrt{(A_{IY}-A_{ZY})^2+(A_{IZ}-A_{ZZ})^2},
\end{align}
and so both of the following hold
\begin{align}
A_{II} &= \cos\left(\frac{\theta_0}{2}\right) = \cos\left(\frac{\theta_1}{2}\right) \nonumber \\
&\approx \cos\left(i \sqrt{(A_{IY}\pm A_{ZY})^2 + (A_{IZ} \pm A_{ZZ})^2}\right).
\end{align}
The Hamiltonian can now be computed via,
\begin{align}
H &= \frac{i\log(U)}{2t},
\end{align}
where $t$ is the duration of each tone.

\section{Hamiltonian error amplifying tomography: $\pm$ rotary single-qubit case}\label{sec:HEAT_sq}

Assume $U$ has the form
\begin{align}
U& =  A_{I}I + A_{Y}Y + A_{Z}Z.
\end{align}
$U$ can be written as
\begin{align}
U &= e^{-i \frac{\theta}{2} \hat{n}\cdot (X,Y,Z)} \nonumber \\
&= \cos\left(\frac{\theta}{2}\right) I - i\sin\left(\frac{\theta}{2}\right) \hat{n}\cdot (X,Y,Z),
\end{align}
and noting $\hat{n}_x=0$,
\begin{align}\label{eq:ntoA_single}
\cos\left(\frac{\theta}{2}\right) &= A_{II},\nonumber \\
\hat{n}_y \sin\left(\frac{\theta}{2}\right) &= i A_Y, \nonumber \\
\hat{n}_z \sin\left(\frac{\theta}{2}\right) &= i A_Z.\nonumber \\
\end{align}

In the small $\theta$ limit we can write,
\begin{align}
\hat{n}_{y} \sin\left(\frac{\theta}{2}\right) &\approx \hat{n}_y \frac{\theta}{2}, \nonumber \\
\hat{n}_{z} \sin\left(\frac{\theta}{2}\right) &\approx \hat{n}_z \frac{\theta}{2}.
\end{align}
Using the sequences in Fig.~\ref{fig:YpZpSeq} we have that for even $N$ and small $\theta$,
\begin{align}\
Y_{\text{err}_N} &= \text{tr}(\rho_N Z) \approx - N\hat{n}_{y}\sin(\theta) \approx -N\hat{n}_{y} \theta, \nonumber \\
Z_{\text{err}_N} &= \text{tr}(\rho_N Y) \approx  N\hat{n}_{z}\sin(\theta) \approx N\hat{n}_{z} \theta,
\end{align}
while for odd $N$ there is mixing between $\hat{n}_{y}$ and $\hat{n}_{z}$. One can therefore estimate the $A$ terms in this limit via,
\begin{align}
A_Y &= -i \hat{n}_y \sin\left(\frac{\theta}{2}\right) \approx \frac{-i \hat{n}_y\theta}{2} \approx \frac{i\text{tr}(\rho_N Z)}{2N}, \nonumber \\
A_Z &= -i \hat{n}_z \sin\left(\frac{\theta}{2}\right) \approx \frac{-i \hat{n}_z\theta}{2} \approx \frac{-i\text{tr}(\rho_N Y)}{2N}.
\end{align}
$A_I$ can be computed by noting that
\begin{align}
\frac{\theta}{2} \approx \pm i \sqrt{A_Y^2+A_Z^2},
\end{align}
and so
\begin{align}
A_I&= \cos\left(\frac{\theta}{2}\right) \approx \cos\left(i \sqrt{A_Y^2+A_Z^2}\right).
\end{align}
The Hamiltonian is given by
\begin{align}
H &= \frac{i\log(U)}{2t},
\end{align}
where $t$ is the duration of each tone.

\section{Purity randomized benchmarking to estimate unitarity}\label{sec:PurityRB}

Purity randomized benchmarking~\cite{Wallman_2015} consists of choosing a gate sequence of length $m$, applying it to the initial state, and measuring the purity $P$ of the output state $\rho$,
\begin{align}
    P=\text{tr}\left(\rho^2\right).
\end{align}
If $\rho$ is written as
\begin{align}
    \rho&= \frac{\mathbbm{1}}{d}+\frac{1}{d}\sum_{j=1}^{d^2-1}n_j\sigma_j,
\end{align}
for the set of $d^2-1$ coefficients $\{n_j\}$ and where $\sigma_j$ is the $j$'th Pauli operator then 
\begin{align}
    n_j &= \langle\sigma_j\rangle=\text{tr}\left(\rho \sigma_j\right).
\end{align}
$P$ is given by
\begin{align}
    P&=\frac{1}{d}+\frac{1}{d}\sum_{j=1}^{d^2-1}n_j^2.
\end{align}

Assuming time-independent, trace-preserving noise $\Eop$ and averaging over many length-$m$ sequences gives the model~\cite{Wallman_2015},
\begin{equation}
\mathbbm{E} [P] = A+B u_\Eop ^{m-1},
\end{equation}
where $u_\Eop \in [0,1]$ is the unitarity of $\Eop$. Let $[\Eop]$ be the Liouville representation of $\Eop$ with respect to the orthonormal Pauli basis, also called the Pauli transfer matrix (PTM) of $\Eop$. The unitarity is given by
\begin{align}
u_\Eop&= \frac{1}{d^2-1}\text{tr}([\Eop_*]^\dagger[\Eop_*]),
\end{align}
where $[\Eop_*]$ is the unital part of $[\Eop]$.

\subsection{One-qubit noise model}

Here we consider $\Eop$ to be the combination of both longitudinal and transverse relaxation. Longitudinal relaxation is modeled via amplitude damping $\Eop_a$ with Kraus operators
\begin{align}
	\Eop_a &:
	\left\{\left[ \begin{array}{cc}
	1	&	0	\\
	0	&	\sqrt{1-\gamma_a}	\\
	\end{array} \right],
	\left[ \begin{array}{cc}
	0	&	\sqrt{\gamma_a}	\\
	0	&	0	\\
	\end{array} \right]\right\},
\end{align}
and transverse relaxaton is modeled via phase damping $\Eop_p$ by the Kraus operators
\begin{align}
	\Eop_p &:
	\left\{\left[ \begin{array}{cc}
	1	&	0	\\
	0	&	\sqrt{1-\gamma_p}	\\
	\end{array} \right],
	\left[ \begin{array}{cc}
	0	&	0	\\
	0	&	\sqrt{\gamma_p}	\\
	\end{array} \right]\right\}.
\end{align}
The PTM's for each are given by
\begin{align}
	R_{\Eop_a} &=
	\left[ \begin{array}{cccc}
	1	&	0	&	0	&	0\\
	0	&	\sqrt{1-\gamma_a}	&	0	&	0\\
	0	&	0	&	\sqrt{1-\gamma_a}	&	0\\
	\gamma_a	&	0	&	0	&	1-\gamma_a\\
	\end{array} \right], \nonumber \\
	R_{\Eop_p} &=
	\left[ \begin{array}{cccc}
	1	&	0	&	0	&	0\\
	0	&	\sqrt{1-\gamma_p}	&	0	&	0\\
	0	&	0	&	\sqrt{1-\gamma_p}	&	0\\
	0	&	0	&	0	&	1\\
	\end{array} \right].
\end{align}
We take $\Eop$ to be the composition of the noise models,
\begin{align}
\Eop &= \Eop_a \circ \Eop_p,
\end{align}
and so since the Liouville representation is multiplicative with respect to channel composition, 
\begin{align}
R_{\Eop} &= R_{\Eop_a}R_{\Eop_p}\nonumber \\
&=	\left[ \begin{array}{cccc}
	1	&	0	&	0	&	0\\
	0	&	\sqrt{1-\gamma_a}\sqrt{1-\gamma_p}	&	0	&	0\\
	0	&	0	&	\sqrt{1-\gamma_a}\sqrt{1-\gamma_p}	&	0\\
	\gamma_a	&	0	&	0	&	1-\gamma_a\\
	\end{array} \right].
\end{align}
The unitarity is then given by
\begin{align}
u_\Eop &= \frac{1}{3}(1-\gamma_a)(3-\gamma_a-2\gamma_p).
\end{align}

The amplitude and phase damping model parameters are given by,
\begin{align}
\gamma_a &= 1-e^{-\Gamma_a t}, \nonumber \\
\gamma_p &= 1-e^{-\Gamma_p t}, \nonumber \\
\end{align}
where
\begin{align}
\Gamma_a &:= \frac{1}{2T_1}, \nonumber \\
\Gamma_p &:= \frac{1}{T_\phi}.
\end{align}
$T_\phi$ is the pure dephasing rate which is related to $T_2$ via
\begin{align}
\frac{1}{T_2} &= \frac{1}{2T_1}+\frac{1}{T_\phi},
\end{align}
so that
\begin{align}
T_\phi &= \frac{2T_1T_2}{2T_1-T_2}.
\end{align}

\subsection{$n$-qubit independent noise model - amplitude and phase damping}

Let us assume an independent noise model on an $n$-qubit system. Each qubit has both amplitude damping, $\Eop_{a,j}$, and phase damping, $\Eop_{p,j}$. The amplitude and phase damping parameters are$\gamma_{a,j}$ and $\gamma_{p,j}$ respectively. The total noise operator is given by
\begin{align}
\Eop= \otimes_{j=1}^n\left(\Eop_{a,j} \circ \Eop_{p,j}\right),
\end{align}
and since the Liouville representation obeys tensor products the PTM is given by,
\begin{align}
R_\Eop &= \otimes_{j=1}^n \left(R_{\Eop_{a,j}}R_{\Eop_{p,j}}\right).
\end{align}

For $n=2$ the unitarity of $\Eop$ is found to be,
\begin{align}
u_\Eop&=1+\frac{1}{15} \Big(\gamma_{a,1}^2 \left(3 \gamma_{a,2}^2+4 \gamma_{a,2}
   (\gamma_{p,2}-2)-4 \gamma_{p,2}+7\right) \nonumber \\
   &+4 \gamma_{a,1} (\gamma_{p,1}-2)
   \left(\gamma_{a,2}^2+\gamma_{a,2} (\gamma_{p,2}-2)-\gamma_{p,2}+2\right)\nonumber \\
   &+\gamma_{a,2}^2 (7-4 \gamma_{p,1})-4 \gamma_{a,2} (\gamma_{p,1}-2) (\gamma_{p,2}-2) \nonumber \\
   &+4 \gamma_{p,1} \gamma_{p,2}-8 \gamma_{p,1}-8
   \gamma_{p,2}\Big).
\end{align}
In the case that the two qubits have the same amplitude damping and phase damping parameters,
\begin{align}
u_\Eop&=\frac{1}{15} (\gamma_a-1) (\gamma_a+2 \gamma_p-3) (\gamma_a (3
   \gamma_a+2 \gamma_p-4) \nonumber \\
   &-2 \gamma_p+5).
\end{align}

To compute $u_\Eop$ for arbitrary $n$, we first note that
\begin{align}\label{eq:unit_sum}
u_\Eop &= \frac{1}{4^n-1}\sum_{i,j=2}^{4^n}R_\Eop(i,j)^2 \nonumber \\
&= \frac{1}{4^n-1}\Big[\text{tr}(R_\Eop^TR_\Eop)-\Big(R_\Eop(1,1)^2\nonumber \\
&+ \sum_{i=2}^{4^n}R_\Eop(i,1)^2 + \sum_{j=2}^{4^n}R_\Eop(1,j)^2\Big)\Big].
\end{align}
The relevant PTM's are given by
\begin{widetext}
\begin{align}\label{eq:Rmatind}
	R_\Eop &=
	\left[ \begin{array}{cccc}
	\otimes_{j=2}^{n} R_{\Eop_j}	&	0	&	0 	&	0	\\
	0	&	\sqrt{1-\gamma_{a,1}}\sqrt{1-\gamma_{p,1}}\otimes_{j=2}^{n} R_{\Eop_j}	&  0	&	0	\\
	0	&	0	&  \sqrt{1-\gamma_{a,1}}\sqrt{1-\gamma_{p,1}}\otimes_{j=2}^{n} R_{\Eop_j}	&	0	\\
	\gamma_{a,1}\otimes_{j=2}^{n} R_{\Eop_j}	&	0	&	 0	&	(1-\gamma_{a,1})\otimes_{j=2}^{n} R_{\Eop_j}
	\end{array} \right], \nonumber \\
	R_\Eop^T &=
	\left[ \begin{array}{cccc}
	\otimes_{j=2}^{n} R_{\Eop_j}^T	&	0	&	0 	&	\gamma_{a,1}\otimes_{j=2}^{n} R_{\Eop_j}^T	\\
	0	&	\sqrt{1-\gamma_{a,1}}\sqrt{1-\gamma_{p,1}}\otimes_{j=2}^{n} R_{\Eop_j}^T	&  0	&	0	\\
	0	&	0	&  \sqrt{1-\gamma_{a,1}}\sqrt{1-\gamma_{p,1}}\otimes_{j=2}^{n} R_{\Eop_j}^T	&	0	\\
	0	&	0	&	 0	&	(1-\gamma_{a,1})\otimes_{j=2}^{n} R_{\Eop_j}^T
	\end{array} \right],
	\end{align}
	\begin{align}
R_\Eop^TR_\Eop &=
 	\left[ \begin{array}{cccc}
	(1+\gamma_{a,1}^2)\otimes_{j=2}^{n} S_j	&	0	&	0 	&	\gamma_{a,1}(1-\gamma_{a,1})\otimes_{j=2}^{n} S_j	\\
	0	&	(1-\gamma_{a,1})(1-\gamma_{p,1})\otimes_{j=2}^{n} S_j	&  0	&	0	\\
	0	&	0	&  (1-\gamma_{a,1})(1-\gamma_{p,1})\otimes_{j=2}^{n} S_j	&	0	\\
	\gamma_{a,1}(1-\gamma_{a,1})\otimes_{j=2}^{n} S_j	&	0	&	 0	&	(1-\gamma_{a,1})^2\otimes_{j=2}^{n} S_j
	\end{array} \right],
\end{align}
\end{widetext}
where $S_j=R_{\Eop_j}^TR_{\Eop_j}$ and so
\begin{align}\label{eq:unit_trace}
\text{tr}(R_\Eop^TR_\Eop) &= \Pi_{j=1}^n\Big(1+\gamma_{a,j}^2+2(1-\gamma_{a,j})(1-\gamma_{p,j})\nonumber \\
&+(1-\gamma_{a,j})^2\Big).
\end{align}
In addition, the non-unital and trace-preserving contributions give
\begin{align}\label{eq:non_unit}
R_\Eop(1,1)^2 + \sum_{i=2}^{4^n}R_\Eop(i,1)^2 &= \Pi_{j=1}^n(1+\gamma_{a,j}^2),\nonumber \\
\sum_{j=2}^{4^n}R_\Eop(1,j)^2 &= 0.
\end{align}
Hence from Eq.'s~\ref{eq:unit_sum},~\ref{eq:unit_trace}, and ~\ref{eq:non_unit}, the unitarity for an independent $n$-qubit noise model is given by,
\begin{align}\label{eq:unit_ind}
u_\Eop &= \frac{1}{4^n-1}\Big[\Pi_{j=1}^n\Big(1+\gamma_{a,j}^2+2(1-\gamma_{a,j})(1-\gamma_{p,j})\nonumber \\
&+(1-\gamma_{a,j})^2\Big) - \Pi_{j=1}^n\left(1+\gamma_{a,j}^2\right)\Big].
\end{align}
Note that if the $n$ qubits have the same amplitude and phase damping parameters,
\begin{gather}
u_\Eop= \frac{1}{4^n-1}\Big[\left(1+\gamma_a^2+2(1-\gamma_a)(1-\gamma_p)+(1-\gamma_a)^2\right)^n \nonumber \\
- (1+\gamma_a^2)^n\Big].
\end{gather}

It is useful to relate $u_\Eop$ more directly in terms of the  $\{u_{\Eop_j}\}$. For $n=2$ we see from $R_\Eop$ in Eq.~\ref{eq:Rmatind}
\begin{align}
    \sum_{i,j=1}^{16} R_\Eop(i,j)^2 &= \sum_{i,j=1}^4R_{\Eop_2}(i,j)^2\Big(1+2(1-\gamma_{a,1})(1-\gamma_{p,1})\nonumber \\
    &+(1-\gamma_{a,1})^2+\gamma_{a,1}^2\Big)\nonumber \\
    &=\left(1+\gamma_{a,2}^2+3u_{\Eop_2}\right)\left(1+\gamma_{a,1}^2+3u_{\Eop_1}\right).
\end{align}
Therefore we have,
\begin{align}\label{eq:unit2ind}
    u_\Eop &= \frac{1}{15}\sum_{i,j=2}^{16}R_\Eop(i,j)^2\nonumber \\
    &=\frac{1}{15}\left[3(1+\gamma_{a,2}^2)u_{\Eop_1}+3(1+\gamma_{a,1}^2)u_{\Eop_2}+9u_{\Eop_1}u_{\Eop_2}\right],
\end{align}
and the non-unital noise prevents writing $u_\Eop$ in terms of solely $u_{\Eop_1}$ and $u_{\Eop_2}$.

\subsection{$n$-independent unital operations}

From the form of the unitarity and Eq.~\ref{eq:unit2ind} we expect that for independent unital operation we can write $u_\Eop$ in terms of just the individual $\{u_{\Eop_j}\}$. Suppose $\Eop$ is the tensor product of $n$ maps $\Eop_j$ where each $\Eop_j$ is a unital quantum operation that acts on the space of $d_j^2 \times d_j^2$ matrices. Hence $d=\Pi_{j=1}^n d_j$ and $\Eop$ is also unital which gives,
\begin{align}
u_\Eop&=\frac{1}{d^2-1}\sum_{m,n=2}^{d^2} R_\Eop (m,n)^2 \nonumber \\
&=\frac{1}{d^2-1}\left(\sum_{m,n=1}^{d^2} R_\Eop(m,n)^2-1\right) \nonumber \\
&=\frac{1}{d^2-1}\left(\text{tr}\left(R_\Eop^TR_\Eop\right)-1\right).
\end{align}
Since the PTM representation is multiplicative with respect to tensor products
\begin{align}
u_\Eop &=\frac{1}{d^2-1}\left(\text{tr}\left(\otimes_{j=1}^n R_{\Eop_j}^TR_{\Eop_j}\right)-1\right)\nonumber \\
&=\frac{1}{d^2-1}\left(\Pi_{j=1}^n \text{tr}\left(R_{\Eop_j}^TR_{\Eop_j}\right)-1\right) \nonumber \\
&=\frac{1}{d^2-1}\left(\Pi_{j=1}^n \left(\frac{d_j^2-1}{d_j^2-1}\right)\text{tr}\left(R_{\Eop_j}^TR_{\Eop_j}\right)-1\right).\nonumber \\
\end{align}
As the $\Eop_j$ are each unital,
\begin{align}
    \text{tr}\left(R_{\Eop_j}^TR_{\Eop_j}\right)&=1+\sum_{m,n=2}^{d_j^2} R_{\Eop_j}(m,n)^2\nonumber \\
    &=1+(d_j^2-1)u_{\Eop_j},
\end{align}
which gives
\begin{align}
u_\Eop &=\frac{1}{d^2-1}\left(\Pi_{j=1}^n (d_j^2-1)\left(\frac{1}{d_j^2-1}+u_{\Eop_j}\right)-1\right)\nonumber \\
&=\frac{1}{d^2-1}\left(\Pi_{j=1}^n\left(1+(d_j^2-1)u_{\Eop_j}\right)-1\right).
\end{align}
Hence in total if $\Eop$ is the tensor product of unital $\Eop_j$ then
\begin{align}
u_\Eop&=\frac{1}{d^2-1}\left(\Pi_{j=1}^n\left(1+(d_j^2-1)u_{\Eop_j}\right)-1\right).
\end{align}

A possible measure of entanglement under the approximation of unital noise is to compute 
\begin{align}
e_\Eop&= u_\Eop-\frac{1}{d^2-1}\left(\Pi_{j=1}^n\left(1+(d_j^2-1)u_{\Eop_j}\right)-1\right),
\end{align}
and for spaces with $d_1=4$ and $d_2=2$ we have
\begin{align}
e_\Eop&= u_\Eop-\frac{1}{63}\left[\left(1+15u_{\Eop_1}\right)\left(1+3u_{\Eop_2}\right)-1\right].
\end{align}

\subsection{Independent $T_1$ and unital noise for $n=2$}

Let us consider an independent noise model consisting of amplitude damping and unital noise on each qubit for the case of $n=2$. Let the amplitude damping and unital noise be denoted by $\Eop_{a,j}$ and $\Lambda_j$ respectively with associated PTM's $R_{\Eop_{a,j}}$ and $R_{\Lambda_j}$. The noise on each qubit is then given by
\begin{align}
\Eop_j &= \Eop_{a,j} \circ \Lambda_j.
\end{align}
First, let us find $u_{\Eop_j}$. As $\Lambda$ is unital, $R_{\Lambda_j}$ is block-diagonal with the $1\times 1$ identity block and a $3 \times 3$ block. Hence 
\begin{align}\label{eq:REopj}
R_{\Eop_j} &=	\left[ \begin{array}{cccc}
	1	&	0	&	0	&	0	\\
	0	&	\:	&  \sqrt{1-\gamma_{a,1}}\left[R_{\Lambda_j}\right]_2	&	\:	\\
	0	&	\:	&  \sqrt{1-\gamma_{a,1}}\left[R_{\Lambda_j}\right]_3	&	\:	\\
	\gamma_{a,1}	&	\:	&	(1-\gamma_{a,1})\left[R_{\Lambda_j}\right]_4	&	\:
	\end{array} \right],
	\end{align}
where $\left[R_{\Lambda_j}\right]_k$ is the $k$'th row of $R_{\Lambda_j}$ and the unitarity is given by
\begin{align}\label{eq:uind}
u_{\Eop_j} &= \frac{1}{3}\Big[(1-\gamma_{a,1})\left[R_{\Lambda_j}\right]_2\left[R_{\Lambda_j}\right]_2^T \nonumber \\
&+ (1-\gamma_{a,1})\left[R_{\Lambda_j}\right]_3\left[R_{\Lambda_j}\right]_3^T \nonumber \\
&+ (1-\gamma_{a,1})^2\left[R_{\Lambda_j}\right]_4\left[R_{\Lambda_j}\right]_4^T\Big].
\end{align}

For $n=2$ we find that
\begin{align}
R_\Eop &=	\left[ \begin{array}{cccc}
	R_{\Eop_2}	&	0	&	0	&	0	\\
	0	&	\:	&  \sqrt{1-\gamma_{a,1}}\left[R_{\Lambda_1}\right]_2 \otimes R_{\Eop_2}	&	\:	\\
	0	&	\:	&  \sqrt{1-\gamma_{a,1}}\left[R_{\Lambda_1}\right]_3 \otimes R_{\Eop_2}	&	\:	\\
	\gamma_{a,1}R_{\Eop_2}	&	\:	&	(1-\gamma_{a,1})\left[R_{\Lambda_1}\right]_4 \otimes R_{\Eop_2}	&	\:
	\end{array} \right],
\end{align}
and so
\begin{align}
    \sum_{i,j=1}^{16}R_\Eop(i,j)^2 &= \sum_{i,j=1}^4 R_{\Eop_2}(i,j)^2 \Big[1+ \gamma_{a,1}^2 \nonumber \\
    &+(1-\gamma_{a,1})\left[R_{\Lambda_j}\right]_2\left[R_{\Lambda_j}\right]_2^T \nonumber \\
&+ (1-\gamma_{a,1})\left[R_{\Lambda_j}\right]_3\left[R_{\Lambda_j}\right]_3^T \nonumber \\
&+ (1-\gamma_{a,1})^2\left[R_{\Lambda_j}\right]_4\left[R_{\Lambda_j}\right]_4^T\Big]\nonumber \\
&= \left(3u_{\Eop_2}+(1+\gamma_{a,2}^2)\right)\left(3u_{\Eop_1}+(1+\gamma_{a,1}^2)\right).
\end{align}
Therefore
\begin{align}
    u_\Eop&=\frac{1}{15}\sum_{i,j=2}^{16} R_\Eop(i,j)^2\nonumber \\
    &=\frac{1}{15}\left[3(1+\gamma_{a,2}^2)u_{\Eop_1}+3(1+\gamma_{a,1}^2)u_{\Eop_2}+9u_{\Eop_1}u_{\Eop_2}\right].
\end{align}

\subsection{$T_1$ and unital noise for control+target and spectator systems}

Here we consider the case of looking at the entanglement across two subsystems; the control+target subsystem of dimension 4 and the spectator system of dimension 2. Here we take the ordering S-C-T so that $d_1=2$ and $d_1=4$. We assume each individual qubit has amplitude damping and each of the two subsystems has arbitrary unital noise maps. Let the unital noise maps be denoted $\Lambda_1$ and $\Lambda_2$, where $\Lambda_1$ acts on the 2-dimensional S space and $\Lambda_2$ acts on the 4-dimensional C+T space. As well, let the amplitude damping maps be denoted $\Eop_{a,j}$ for $j=1,2,3$. Hence the noise models on S and C+T are given by $\Eop_1=\Eop_{a,1} \circ \Lambda_1$ and $\Eop_2=\left(\Eop_{a,2} \otimes \Eop_{a,3}\right) \circ \Lambda_2$ respectively. 

First note that Eq.~\ref{eq:uind} gives $u_{\Eop_1}$. Next, since $R_{\Eop_{a,2} \otimes \Eop_{a,3}}$ takes the form
\begin{gather}
	\left[ \begin{array}{cccc}
	R_{\Eop_{a,3}}	&	0	&	0 	&	0	\\
	0	&	K_2R_{\Eop_{a,3}}	&  0	&	0	\\
	0	&	0	&  K_2R_{\Eop_{a,3}}	&	0	\\
	\gamma_{a,1}R_{\Eop_{a,3}}	&	0	&	 0	&	K_2^2 R_{\Eop_{a,3}}
	\end{array} \right],
\end{gather}
for $K_2=\sqrt{1-\gamma_{a,2}}$ we have
\begin{align}\label{eq:Reop2}
	R_{\Eop_2}&=R_{\Eop_{a,2} \otimes \Eop_{a,3}}R_{\Lambda_2} \nonumber \\
	&=
	\left[ \begin{array}{c}
	R_{\Eop_{a,3}}\left[R_{\Lambda_2}\right]_1^4	\\
	\sqrt{1-\gamma_{a,2}}R_{\Eop_{a,3}}\left[R_{\Lambda_2}\right]_5^8	\\
	\sqrt{1-\gamma_{a,2}}R_{\Eop_{a,3}}\left[R_{\Lambda_2}\right]_9^{12}	\\
	\gamma_{a,1}R_{\Eop_{a,3}}\left[R_{\Lambda_2}\right]_1^4	+ (1-\gamma_{a,2})R_{\Eop_{a,3}}\left[R_{\Lambda_2}\right]_{13}^{16}
	\end{array} \right],
\end{align}
where $\left[\Lambda_2\right]_i^j$ is the matrix corresponding to the $i$'th through $j$'th rows of $[\Lambda_2]$. Therefore
\begin{align}
u_{\Eop_2} &= \frac{1}{63}\sum_{i,j=2}^{64} R_{\Eop_2}(i,j)^2,
\end{align}
where $R_{\Eop_2}$ is given in Eq.~\ref{eq:Reop2}. An explicit calculation of $u_{\Eop_2}$ will not be required to calculate $u_\Eop$.

The PTM of the composite system is given by $R_{\Eop}=R_{\Eop_1}\otimes R_{\Eop_2}$ where $R_{\Eop_1}$ is given by Eq.~\ref{eq:REopj} for $j=1$ and $R_{\Eop_2}$ is given as above. Thus,
\begin{align}
R_\Eop &= R_{\Eop_1}\otimes R_{\Eop_2} \nonumber \\
&=\left[ \begin{array}{cccc}
	R_{\Eop_2}	&	0	&	0	&	0	\\
	0	&	\:	&  \sqrt{1-\gamma_{a,1}}\left[R_{\Lambda_1}\right]_2 \otimes R_{\Eop_2}	&	\:	\\
	0	&	\:	&  \sqrt{1-\gamma_{a,1}}\left[R_{\Lambda_1}\right]_3 \otimes R_{\Eop_2}	&	\:	\\
	\gamma_{a,1} R_{\Eop_2} 	&	\:	&	(1-\gamma_{a,1})\left[R_{\Lambda_1}\right]_4 \otimes R_{\Eop_2}	&	\:
	\end{array} \right],
\end{align}
and so
\begin{gather}
\sum_{i,j=1}^{64} R_\Eop(i,j)^2 = \text{tr}\left(R_{\Eop_2} R_{\Eop_2}^T\right) + \gamma_{a,1}^2\text{tr}\left(R_{\Eop_2} R_{\Eop_2}^T\right)\nonumber \\
+ (1-\gamma_{a,1})\text{tr}\left(\left([R_{\Lambda_1}]_2\otimes R_{\Eop_2}\right) \left([R_{\Lambda_1}]_2\otimes R_{\Eop_2}\right)^T\right) \nonumber \\
+ (1-\gamma_{a,1})\text{tr}\left(\left([R_{\Lambda_1}]_3\otimes R_{\Eop_2}\right) \left([R_{\Lambda_1}]_3\otimes R_{\Eop_2}\right)^T\right) \nonumber \\
+ (1-\gamma_{a,1})^2\text{tr}\left(\left([R_{\Lambda_1}]_4\otimes R_{\Eop_2}\right) \left([R_{\Lambda_1}]_4\otimes R_{\Eop_2}\right)^T\right) \nonumber \\
=\text{tr}\left(R_{\Eop_2} R_{\Eop_2}^T\right)\Big[1+\gamma_{a,1}^2 + (1-\gamma_{a,1})\Big[ \text{tr}\left([R_{\Lambda_1}]_2 [R_{\Lambda_1}]_2^T \right)\nonumber \\
+ \text{tr}\left([R_{\Lambda_1}]_3 [R_{\Lambda_1}]_3^T \right)\Big]  + (1-\gamma_{a,1})^2 \text{tr}\left([R_{\Lambda_1}]_4 [R_{\Lambda_1}]_4^T \right)\Big] \nonumber \\
=\text{tr}\left(R_{\Eop_2} R_{\Eop_2}^T\right)\left[1+\gamma_{a,1}^2 + 3u_{\Eop_1}\right],
\end{gather}
where the last line follows from Eq.~\ref{eq:uind}. Now it is straightforward to verify that the first column of $R_{\Eop_2}$ in Eq.~\ref{eq:Reop2} has contribution
\begin{align}
\sum_{i=1}^{64}R_{\Eop_2}(i,1)^2 &= 1+\gamma_{a,2}^2+\gamma_{a,2}^2\gamma_{a,3}^2+\gamma_{a,3}^2\nonumber \\
&=(1+\gamma_{a,2})^2(1+\gamma_{a,3}^2),
\end{align}
so that
\begin{align}
\text{tr}\left(R_{\Eop_2} R_{\Eop_2}^T\right) &= 15 u_{\Eop_2} + (1+\gamma_{a,2})^2(1+\gamma_{a,3}^2),
\end{align}
which gives
\begin{gather}
\sum_{i,j=1}^{64} R_\Eop(i,j)^2 \\
= \left(15 u_{\Eop_2} + (1+\gamma_{a,2}^2)(1+\gamma_{a,3}^2)\right)\left[1+\gamma_{a,1}^2 + 3u_{\Eop_1}\right].
\end{gather}
Next, note that
\begin{align}
\sum_{i=1}^{64} R_\Eop(i,1)^2 &= (1+\gamma_{a,1}^2)\sum_{i=1}^{64}R_{\Eop_2}(i,1)^2 \nonumber \\
&= (1+\gamma_{a,1}^2)(1+\gamma_{a,2}^2)(1+\gamma_{a,3}^2),
\end{align}
which gives $u_\Eop$ in terms of $u_{\Eop_1}$, $u_{\Eop_2}$, and the decay parameters $\{\gamma_{a,j}\}$,
\begin{gather}\label{eq:ueopT1}
u_\Eop = \frac{1}{63}\left( \sum_{i,j=1}^{64} R_\Eop(i,j)^2 - \sum_{i=1}^{64} R_\Eop(i,1)^2\right) \nonumber \\
= \frac{1}{63}\Big[\left(15 u_{\Eop_2} + (1+\gamma_{a,2}^2)(1+\gamma_{a,3}^2)\right)\left(1+\gamma_{a,1}^2 + 3u_{\Eop_1}\right)\nonumber \\
- (1+\gamma_{a,1}^2)(1+\gamma_{a,2}^2)(1+\gamma_{a,3}^2)\Big] \nonumber \\
= \frac{1}{63}\Big[ 15 u_{\Eop_2}\left(1+\gamma_{a,1}^2\right) + 45u_{\Eop_1}u_{\Eop_2} \nonumber \\
+ 3u_{\Eop_1}(1+\gamma_{a,2}^2)(1+\gamma_{a,3}^2)\Big].
\end{gather}
Re-ordering to C-T-S gives,
\begin{gather}
    u_\Eop = \frac{1}{63}\Big[ 15 u_{\Eop_1}\left(1+\gamma_{a,3}^2\right) + 45u_{\Eop_1}u_{\Eop_2} \nonumber \\
+ 3u_{\Eop_2}(1+\gamma_{a,1}^2)(1+\gamma_{a,2}^2)\Big].
\end{gather}

\section{Numerical Methods}\label{sec:Numerics}

While first principles simulations readily reveal the qualitative behavior of target rotary pulsing, we leave a full numerical study of the effect of target rotary pulsing for future work. Here we show that a simple numerical model captures the data quantitatively with relatively few fit parameters that are reasonable in light of our theoretical understanding of pulsed, coupled transmon qubits.  To estimate the expected error of our echoed pulse sequence, we simulate the HEAT pulse sequences directly in the computational subspace of the three-qubit system, allowing anharmonicity to affect certain terms of our Hamiltonian within the computational subspace, but leakage is not included (and in fact does not play a large role in the parameter space explored in this work, as we have checked both experimentally and numerically). In our numerical fits, we substitute
\begin{equation}
U_{\pm}=e^{i\sum_{i,j,k\in X,Y,Z}\theta^{ijk}_{\pm}\sigma_{ijk}}
\end{equation}
for the 
evolution under positive/negative  ${\rm CR}_\pm\frac{\pi}{4}$ and ${\rm R}_\pm$  in Figs. 1 and 2 in the main text.  Here we adopt the convention used in the main text that
\begin{equation}
\sigma_{ijk}=\sigma_i\otimes\sigma_j\otimes\sigma_k
\end{equation}
where $\sigma_i$ acts on the control Q11 Hilbert space, $\sigma_j$ acts on the target Q12, and $\sigma_k$ acts on the target-spectator Q13.
$|\theta^{ijk}_{+}|=|\theta^{ijk}_{-}|$, with  $\theta^{ZXI}$, $\theta^{IXZ}$, $\theta^{IXI}$, $\theta^{ZYI}$, $\theta^{IYZ}$, and $\theta^{IYI}$ changing signs from $U_+$ to $U_-$ while $\theta^{ZZI}$, $\theta^{IZZ}$, and $\theta^{IZI}$ do not. 
Only these terms are considered in our model Hamiltonian, all of which are expected from previous CR perturbation theory work and a simple model of crosstalk \cite{Magesan2020,Sheldon2016}.  While a more inclusive model might cover errors from pulsing imperfections ($\theta^{XII}$, $\theta^{YII}$, etc.), we limit the discussion here and revel in the rich physics resulting from only the largest of the Hamiltonian error terms.\\

We model numerically only the relevent two- or three-qubit subsystem of the 20-qubit  \emph{$ibmq\_johannesburg$} device.   
Numerical fits to Fig. ~\ref{figure:Fig1} are two-qubit computations with $\theta^{IZZ}_1=\theta^{IXZ}_1=\theta^{IYZ}_1=0$, justified by the fact that the target spectator Q13 remains in its ground state for this experiment and its preceding calibrations.   In other words, while qubits Q11 and Q12 are calibrated `simultaneously' for Fig. ~\ref{figure:Fig1} data, they are `isolated' from all spectators which are in their ground state with high probability \cite{McKay2020}. Target spectator Q13 is included in three-qubit simulations of Figs. ~\ref{figure:Fig2} and ~\ref{figure:Fig3} because it is not only calibrated simultaneously with the control and target qubits but is an integral part of the presented experiments.  By comparison, target spectator Q7 is not included in any numerical modeling as it is calibrated in isolation and is not used in any experiment.\\

Focusing on Fig. ~\ref{figure:Fig1}, each of the nonzero angles are fit to quadratic functions of the applied rotary amplitude $x$.  The values of each parameter producing the numeric curve of Fig. ~\ref{figure:Fig1} are shown in Table. ~\ref{table:Fig1Num}. The terms in bold are numerically optimized using a least squared fit of the experimental data. $\theta^{ZZI}_{0}$ is fixed by an independent measurement of the ZZ rate (over 2) times the pulse time, $206.22$ ns.

\begin{equation}
\theta^{ijk}(x)=\theta^{ijk}_{0}+\theta^{ijk}_{1}\cdot x+\theta^{ijk}_{2}\cdot x^2
\end{equation}
\begin{table}[ht!]
\begin{center}
 \begin{tabular}{|p{0.1\textwidth}||p{0.11\textwidth}|p{0.11\textwidth}|p{0.11\textwidth}|}
 \hline
 ijk
  &$\theta^{ijk}_{0}$  & $\theta^{ijk}_{1}$ & $\theta^{ijk}_{2}$ \\ 
 \hline
 IXI & \textbf{-4.63e-1} & ~1.00 & ~0.00 \\ 
\hline
IYI &~\textbf{7.98e-3} &~0.-- & ~0.00 \\ 
\hline
IZI & ~\textbf{2.55e-2} & \textbf{-1.23e-3 }& \textbf{-2.69e-3} \\ 
\hline
ZZI &-1.59e-2 & ~0.00 &\textbf{ 2.11e-3}  \\ 
\hline
ZYI & ~\textbf{7.05e-3 }&~0.00 &~0.00 \\ 
\hline
ZXI &~~~~~$\frac{\pi}{8}$&~0.00& ~0.00 \\ 
\hline
\end{tabular}
\end{center}
\caption{Parameters in numerical model used to describe Fig.  ~\ref{figure:Fig1}, bold parameters were fit to data.}
\label{table:Fig1Num}
\end{table}
To fit Fig. ~\ref{figure:Fig2}, we fix many parameters from our fit to Fig. ~\ref{figure:Fig1} and add terms that describe the coupling between the target and target-spectator, as shown in Table ~\ref{table:Fig2Num}. \\

\begin{table}[ht!]
\begin{center}
 \begin{tabular}{|p{0.1\textwidth}||p{0.11\textwidth}|p{0.11\textwidth}|p{0.11\textwidth}|}
 \hline
 ijk
  & $\theta^{ijk}_{0}$  & $\theta^{ijk}_{1}$ & $\theta^{ijk}_{2}$ \\ 
 \hline
 IXI & \textbf{-2.97e-1} & ~1.00 & ~0.00 \\ 
\hline
IYI &~0.00 &~0.00 & ~0.00 \\ 
\hline
IZI & ~\textbf{4.88e-2} & {-1.23e-2}& \textbf{-5.84e-3} \\ 
\hline
ZZI &-1.59e-2 & ~0.00 &\textbf{ 2.11e-3}  \\ 
\hline
IZZ &-2.66e-2 & ~0.00 & \textbf{-3.02e-3} \\
\hline
IZY & ~0.00&~0.00 &~0.00 \\ 
\hline
ZYI & ~0.00&~0.00 &~0.00 \\ 
\hline
ZXI & ~0.00 &~0.00& ~0.00 \\ 
\hline
IYZ & ~0.00&  ~0.00 &~0.00 \\ 
\hline 
IXZ & ~\textbf{7.21e-2}&  \textbf{-1.85e-2} &~0.00 \\ 
\hline
\end{tabular}
\end{center}
\caption{Parameters in numerical model used to describe Fig. ~\ref{figure:Fig2}, bold parameters were fit to data.}
\label{table:Fig2Num}
\end{table}

In Fig. ~\ref{figure:Fig3}, the unitary evolution of the three qubit system under an entangling pulse $U_{\rm ent}\simeq e^{i\frac{\pi}{4}\sigma_{ZXI}}$ is calculated using the fit for $\theta^{ZXI}_{0}$, $\theta^{IXI}_{0}$,
$\theta^{ZZI}_{0}$,
$\theta^{IZI}_{0}$,
$\theta^{ZYI}_{0}$, and
$\theta^{IYI}_{0}$
obtained in the fit for Fig. ~\ref{figure:Fig1} and
$\theta^{IZZ}_{0}$, $\theta^{IZY}_{0}$,
$\theta^{IYZ}_{0}$,
$\theta^{IXZ}_{0}$,
obtained in the fit for Fig. ~\ref{figure:Fig2}.
The unitarity of the 2Q and 1Q subsystems are then calculated after the performance of $U_{\rm ent}$, as well as the entropy of entanglement between these two subsystems.

\bigskip
\section{Backend details and Quantum Volume comparison}\label{sec:Volume}

\begin{widetext}
\begin{figure*}
\centering
\includegraphics[width=0.9\textwidth]{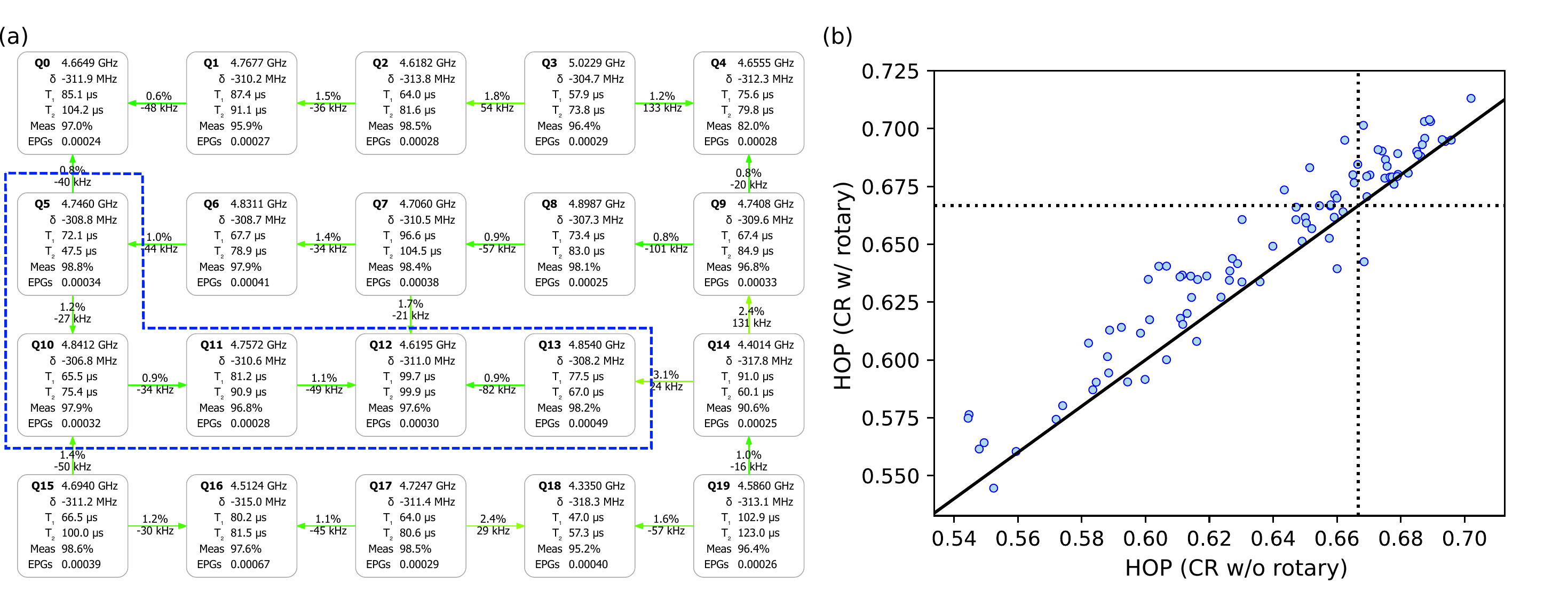}
\caption{\label{fig:QV} (a) Typical parameters and performance of \emph{$ibmq\_johannesburg$}. QV32 HOP on subset A (blue dashed line) with and without rotary is presented in Table I in the main text. (b) A comparison of HOP, using all possible 5 qubit linear subsets on \emph{$ibmq\_johannesburg$}, between the standard echoed CR and CR with rotary.} 
\label{figure:QVfig}
\end{figure*}
\end{widetext}

 In Fig. ~\ref{figure:QVfig}(a) we provide a representative performance map of \emph{$ibmq\_johannesburg$}. Experimental data presented in this paper was taken over several days and so is subject to parameter drift. 
 For the average QV32 HOPs with and without target rotary (presented in Table I of the main text), we tested all possible 5 qubit linear subsets of \emph{$ibmq\_johannesburg$}. For this experiment, the frequencies of the qubits in each subset were first calibrated jointly to account for average static ZZ and then two-qubit gates were optimized for each echoed CR condition prior to executing 150 optimized QV32 circuits. In Fig. ~\ref{figure:QVfig}(b), we plot the HOP for each of the calibration conditions for each subset. Statistically, we see HOP from most subsets fall above the black guide-line (denoting equal HOP in both conditions) and infer the performance improvement added by target rotary. As only 150 circuits were performed to reduce run-time, we note that these measurements do not meet the confidence threshold to confirm QV32. As a proof of concept, we selected subset A (denoted in Fig. ~\ref{figure:QVfig}(a) with a blue dashed line) and executed the necessary number of circuits to confirm that with target rotary the HOP passes the QV32 threshold and is unable to pass without target rotary. 

\bigskip

\end{document}